\documentclass[preprint, 12pt, 3p]{elsarticle}

\usepackage{graphicx}

\usepackage[utf8]{inputenc}
\usepackage{hyperref}
\usepackage[usenames,dvipsnames]{xcolor}
\usepackage{amsmath}
\usepackage{mathrsfs}
\usepackage{amsfonts}
\usepackage{amssymb}
\usepackage{accents}
\usepackage{amsthm}
\usepackage{subcaption}
\usepackage{float}
\usepackage{appendix} 

\newcommand{\Tufd}[1]{{\mathcal{T}}_\delta{\left ({#1}\right )}} 
\def\1mY{\frac{1}{|Y|}}
\def\Ab{{\bmi{A}}}
\def\Acal{\mathcal{A}}
\def\Acalbf{{\mbox{\boldmath$\Acal$\unboldmath}}}
\def\Aop{{{\rm A} \kern-0.6em{\rm A}}}%
\newcommand\Appx[1]{Appendix~\ref{#1}}
\def\Bb{{\bmi{B}}}
\def\Bcal{\mathcal{B}}
\def\Bcalbf{{\mbox{\boldmath$\Bcal$\unboldmath}}}

\def\DDcalbf{{\mbox{{\boldmath$\Ical$\unboldmath}\kern-0.5em{\boldmath$\Dcal$\unboldmath}}}}

\def\Dcal{\mathcal{D}}
\def\Dop{{{\rm I} \kern-0.2em{\rm D}}}

\newcommand\HOpdb{{\bf{H}}_{\#0}^1}

\def\Hbm{{\bf{H}}}
\def\Hcal{\mathcal{H}}
\def\Hcalbf{{\mbox{\boldmath$\mathcal{H}$\unboldmath}}}

\def\Hdb{{\bf{H}}^1}
\def\Hop{{{\rm I} \kern-0.2em{\rm H}}}%
\def\Hpdb{{\bf{H}}_\#^1}
\def\Hpdbav{\tilde{\Hbm}_\#^1}
\def\Ib{{\bmi{I}}}
\def\Ical{\mathcal{I}}
\def\Kb{{\bmi{K}}}

\def\Kcal{\mathcal{K}}
\def\Kcalbf{{\mbox{\boldmath$\Kcal$\unboldmath}}}
\def\Lb{{\bf{L}}}

\def\Mcal{\mathcal{M}}

\newcommand\MeanY[1]{\Mcal_{Y}{\left (#1\right )}}
\newcommand\MeanYd[2]{\Mcal_{Y_{#1}}{\left (#2\right )}}

\def\Om{\Omega}
\def\Pcal{\mathcal{P}}
\def\Pcalbf{{\mbox{\boldmath$\mathcal{P}$\unboldmath}}}
\def\Pibf{{\mbox{\boldmath$\Pi$\unboldmath}}}

\def\Qcal{\mathcal{Q}}
\def\Qcalbf{{\mbox{\boldmath$\Qcal$\unboldmath}}}
\def\R{\hbox{\rm I\kern-0.2em R}}
\def\RR{{\mathbb{R}}}

\def\Rcal{\mathcal{R}}
\def\Rcalbf{{\mbox{\boldmath$\Rcal$\unboldmath}}}

\def\Scal{\mathcal{S}}
\def\Scalbf{{\mbox{\boldmath$\mathcal{S}$\unboldmath}}}
\def\Tcal{\mathcal{T}}

\def\Tuf#1{{\mathcal{T}}_\veps{\left ({#1}\right )}} 
\newcommand\Tuftxt{\Tcal_\veps\,}

\def\Vb{{\bmi{V}}}
\def\Wb{{\bmi{W}}}

\def\Z{\hbox{\rm Z\kern-0.3em Z}}
\def\ZZ{{\mathbb{Z}}}

\def\aYm#1#2{a_m \left ({#1},\,{#2}\right )}
\def\aYs#1#2{a_s \left ({#1},\,{#2}\right )}
\def\ab{{\bmi{a}}}
\def\bYm#1#2{b_m \left ({#1},\,{#2}\right )}
\def\bb{{\bmi{b}}}

\newcommand\blk{{\rm{blk}}} 

\def\bmi#1{\textbf{\textit{#1}}}
\def\cYm#1#2{c_m \left ({#1},\,{#2}\right )}

\newcommand\cf{{{cf.~}}}
\newcommand\chE[1]{{#1}}
\def\chibf{{\mbox{\boldmath$\chi$\unboldmath}}}
\def\chii{\mbox{$\chi$ \kern-1.0em $=$}}
\newcommand\cx{{\rm{[c]}}}

\newcommand\dSy{\mathrm{\,dS}_{y}}  %
\newcommand\dlt{{\delta}}

\def\dvg{\mbox{\rm div}}
\def\eb{{\bmi{e}}}

\newcommand\eeb[1]{\eb({#1})}
\def\eebx#1{\eb_x({#1})}
\def\eeby#1{\eb_y({#1})}
\newcommand\eebz[1]{\eb_z({#1})}
\newcommand\eg{{\it{e.g.~}}}

\newcommand\epsdlt{{\veps,\delta}}
\def\etal{{et~al.}}
\newcommand\ext{{\rm ext}}
\def\fb{{\bmi{f}}}

\newcommand\flow{{\rm{flow}}}
\def\fsi{{\rm{fs}}}
\newcommand\fx{{\rm{[f]}}}
\def\gb{{\bmi{g}}}
\def\hb{{\bmi{h}}}
\newcommand\ie{{\it{i.e.~}}}

\def\intY{\:\sim \kern-1.17em \int}
\def\intYs{- \kern-0.82em \int }
\def\intYsmall{{\small \sim} \kern-0.95em \int}
\def\ipZc#1#2{\left \langle{#1},\,{#2}\right \rangle_{Z_c}}
\def\jb{{\bmi{j}}}

\newcommand\lhs{{{\rm l.h.s.~}}}
\def\linx#1#2{\kern 0.1em _{#1}\kern-0.05em {#2}}

\def\luinx#1#2#3{\kern 0.1em _{#1}\kern-0.22em ^{#2}\kern-0.17em {#3}}

\newcommand\meso{{\rm mes}}
\newcommand\mic{{\rm mic}}
\newcommand\mx{{\rm{[m]}}}
\def\nb{{\bmi{n}}}

\newcommand\notilde[1]{\what{#1}}

\def\omegabf{{\mbox{\boldmath$\omega$\unboldmath}}}

\newcommand\oneb{\bmi{1}}
\def\pd{\partial}

\newcommand\por{{\rm{por}}}

\def\psibf{{\mbox{\boldmath$\psi$\unboldmath}}}

\newcommand\rhs{{{\rm r.h.s.~}}}

\def\sigmabf{{\mbox{\boldmath$\sigma$\unboldmath}}}
\newcommand\sx{{\rm{[s]}}}
\def\tb{{\bmi{t}}}
\def\thetabf{{\mbox{\boldmath$\theta$\unboldmath}}}
\def\ub{{\bmi{u}}}

\def\uinx#1#2{\kern 0.1em ^{#1}\kern-0.15em {#2}}

\def\vb{{\bmi{v}}}

\def\veps{\varepsilon}
\def\vepsdel{{\varepsilon\delta}}
\def\vphi{\varphi}
\def\vtheta{\vartheta}
\def\vthetabf{{\mbox{\boldmath$\vartheta$\unboldmath}}}
\def\wb{{\bmi{w}}}

\def\what#1{\widehat{#1}}
\newcommand\wrt{{\rm with respect to~}}

\def\xibf{{\mbox{\boldmath$\xi$\unboldmath}}}

\def\zb{{\bmi{z}}}

\newcommand{\eq}[1]{(\ref{#1})}
\newcommand{\wtilde}[1]{\widetilde{#1}}
\newcommand{\ol}[1]{\overline{#1}}
\newcommand{\cwto}{\rightharpoonup}

\newcounter{remark}
\newenvironment{myremark}[1]%
{\vspace{7pt} \noindent{\bf Remark
  \refstepcounter{remark}\theremark. \protect\label{#1} \hspace{-2mm}}\rm }%
{\vspace*{-7pt} \flushright $\triangle$\\ } 

\newcounter{proposition}
\newenvironment{myproposition}[1]%
{\vspace{7pt} \noindent{\bf Proposition
  \refstepcounter{proposition}\theproposition. \protect\label{#1} \hspace{-2mm}}\rm }%
{\vspace*{-7pt} \flushright $\triangle$\\ } 

\newenvironment{myproof}[1]%
{\vspace{7pt} \noindent{\bf Proof} of {#1} \hspace{-2mm}\rm }%
{\vspace*{-7pt} \flushright $\square$\\ } 

\begin{document}

\begin{frontmatter}

\title{The Biot-Darcy-Brinkman model of flow in deformable double porous media; homogenization and numerical modelling}

\author[1]{Eduard Rohan\corref{cor1}}
\ead{rohan@kme.zcu.cz}

\author[1]{Jana Turjanicov\'a}
\ead{turjani@students.zcu.cz}

\author[1]{Vladim\'{\i}r Luke\v{s}}
\ead{vlukes@kme.zcu.cz}

\address[1]{European Centre of Excellence, NTIS -- New Technologies for
Information Society, Faculty of Applied Sciences, University of West Bohemia,
Univerzitni\'{\i} 22, 30614 Pilsen, Czech Republic}

\cortext[cor1]{I am corresponding author}

\begin{abstract}
  In this paper we present the two-level homogenization of the flow in a deformable double-porous structure described at two characteristic scales. The higher level porosity associated with the mesoscopic structure is constituted by channels in a matrix made of a microporous material consisting of elastic skeleton and pores saturated by a viscous fluid. The macroscopic model is derived by the homogenization of the flow in the heterogeneous structure characterized by two small parameters involved in the two-level asymptotic analysis, whereby a  scaling ansatz is adopted to respect the pore size differences. The first level upscaling of the fluid-structure interaction problem yields a  Biot continuum describing the mesoscopic matrix coupled with the Stokes flow in the channels. The second step of the homogenization leads to a macroscopic model involving three equations for displacements, the mesoscopic flow velocity and the micropore pressure. Due to interactions between the two porosities, the macroscopic flow is governed by a Darcy-Brinkman model comprising two equations which are coupled with the overall equilibrium equation respecting the hierarchical structure of the two-phase medium.   Expressions of the effective macroscopic parameters of the homogenized double-porosity continuum are derived, depending on the characteristic responses of the mesoscopic structure. Some symmetry and reciprocity relationships are shown and issues of boundary conditions are discussed. The model has been implemented in the finite element code SfePy which is well-suited for computational homogenization. A numerical example of solving a nonstationary problem using mixed finite element method is included.
\end{abstract}  


\end{frontmatter}

\section{Introduction}
The fluid-saturated porous media (FSPM) present an important field of contemporary research in multiscale mathematical modelling and computational methods. Though the topic has already been studied in the last century and the seminal works of M.A.~Biot  \cite{Biot1941,Biot1955} are still relevant, some issues pertain to be a challenge for development of adequate and accurate models and efficient computational methods respecting hierarchical arrangement of the structure. During recent decades, various models and modelling approaches have been proposed with  perspectives of their applications in geosciences, material engineering, including applications in the tissue biomechanics. In many kinds of materials, such as rocks or biological tissues, the  poroelasticity can be characterized at several scales \cite{Gailani2011,rohan-etal-jmps2012-bone}. In particular, namely the double-porosity media are of interest. They consist of two very distinct porous systems so that their interaction has a strong influence on the fluid transfer and other mechanical properties when dealing with deformable structures.  In general, the ``primary'' and the ``dual'' porosities can be distinguished. These two systems characterized by very different pore sizes are arranged hierarchically, one is embedded in the other.

The double porosity concept was first introduced in the context of geomechanics by Barenblatt \etal \cite{Barenblatt1960} who studied flows in cracked rocks. In this approach, one of the two porous systems is associated with the cracks and the other one with the porous matrix, see \cite{Berryman2002,Berryman2012}. Therein the key hypothesis in the modeling consists at defining two averaged fluid pressures at any spatial position, each associated with one porous systems.
There is also another conception to introduce  ``double-porosity'' media as strongly heterogeneous porous structures. In this context the double porosity presents two co-existing systems of connected porosities with large differences in the permeability coefficients. Therefore, such media are sometimes called ``double permeability'' media, to distinguish them form the ``nested porous structures'' described above. The scaling ansatz introduced by Arbogast \cite{Arbogast1990} in his work related to flow in a rigid double porous medium has been pursued \eg in \cite{rohan-etal-jmps2012-bone} to deal with deformable structures.

The homogenization theory  which developed from 1970s, has contributed significantly in the comprehension of the macroscopic behavior of these
materials \cite{Auriault1977,Auriault1992,Auriault1993}, \cf \cite{Royer-Auriault-Boutin-1996}, where the two-level upscaling was employed to treat double-porosity reservoirs.
The homogenization technique based on the
asymptotic analysis with respect to the scale parameter \cite{Bensoussan1978book,Sanchez1980Book,Cioranescu-etal-2008} presents a very rigorous way of upscaling thermodynamic systems described by boundary value problems involving partial differential equations and boundary conditions. Besides the formal asymptotic expansion method,   the two-scale convergence \cite{Allaire1992} and the \emph{periodic unfolding} method of homogenization \citep{Cioranescu-etal-2008} can be used to study rigorously such systems by means of the functional analysis.

In this paper we consider the flow in a deformable double-porous structure described at two characteristic scales. The higher level porosity associated with the mesoscopic structure is constituted by channels in a matrix made of a microporous material consisting of elastic skeleton and pores saturated by a viscous fluid. For upscaling we pursue the two-level homogenization approach applied recently in \cite{Rohan-etal-CaS2017} to derive the macroscopic model (called the Darcy-Brinkman model, \cf \cite{Brinkman1947,Brinkman1949,Auriault-validity-Brinkman2009,Brown2015}) describing viscous flows in rigid double porous structures. In that work, a thorough discussion of various models and upscaling methods was presented, such as volume averaging proposed by Ochoa-Tapia and Whitaker \cite{OchoaTapia19952635,OCHOATAPIA19952647}. Those readers interested in a commented survey of relevant works dealing with upscaling the nested porous structures are encouraged to consult \cite{Rohan-etal-CaS2017}. The 1st level of the homogenization of a rigid porous structure was studied in \cite{CDG-Stokes-2005}. Therein the asymptotic analysis of the Stokes flow with scale dependent viscosity in the micropores which are drained into the mesoscopic channels leads to a mesoscopic model combining the Stokes and Darcy flow models. As a by-product, modified transmission
conditions of the Saffman type \cite{Saffman1971}, cf. \cite{Jaeger-Mikelic-2009,Lesinigo-etal-multi-Brinkman2011}  were obtained automatically on interfaces between the upscaled microporosity and the mesoscopic channels.

\paragraph{The main contribution of the present paper} The two-level homogenized model derived in \cite{Rohan-etal-CaS2017} is extended here to account for deformations of the microporous skeleton. To this aim we pursue the general conception of the reiterated homogenization presented in \cite{trucu2012three}  to upscale elliptic problems governed by the Laplace operator. In our work, the hierarchical  structure is described by two independent scale parameters which are subject of the subsequent asymptotic analyses of the fluid-structure interaction problem. Up to our knowledge, such a treatment has not been considered so far in the published literature. 

The first level homogenization of the fluid-structure interaction in the microporosity 
yields the mesocopic model describing the Biot-Darcy medium in the matrix interacting with the Stokes flow in the mesoscopic channels. The homogenized coefficients comprise the classical poroelasticty coefficients and static permeability, \cf \cite{rohan-etal-CMAT2015}.
Then the second level homogenization is performed to derive the macroscopic model involving three equations for displacements, the mesoscopic flow velocity and the micropore pressure. The macroscopic flow is governed by a Darcy-Brinkman model comprising two equations which are coupled with the overall equilibrium equation respecting the hierarchical porous structure of the two-phase medium. Although the limit two-scale equations are derived for a compressible fluid, the macroscopic homogenized model defined in terms of homogenized coefficients  is obtained under the restriction to incompressible fluids. The general case brings about complications in splitting the scales in the second level homogenization, since the time dependence is more involved in the two-scale problem. Nevertheless, extension of the present model for compressible fluids and inertia effects  will be treated in our future research.

\paragraph{The outline of the paper is as follows} In Section~\ref{sec-FSImic}, we introduce the fluid-structure interaction problem in the double porous structure featured by two scale parameters, $\veps$ and $\dlt$, characterizing the scale of the micro- and meso-scopic porosities. Section~\ref{sec-homog} is devoted to the homogenization procedure. The 1st level upscaling procedure associated with the model asymptotics for $\veps\rightarrow 0$ is explained in Section~\ref{sec-1-homog}. The resulting Biot model coupled with the Stokes flow in the mesoscopic channels is defined in terms of the standard poroelastic coefficients. In Section~\ref{sec-2-homog}, the subsequent  asymptotic analysis for $\dlt\rightarrow 0$ is pursued. The local and global two-scale limit equations are obtained using the convergence result. Then, in Section~\ref{sec-mac-i}, the macroscopic homogenized model is derived for the incompressible fluid. Although the model is obtained for a particular system of the boundary conditions, a generalization for other combinations of the boundary conditions is discussed. Finally, in Section~\ref{sec-numex}, we report an example which illustrates some interesting features of the upscaled model describing the flows in the double porosity elastic structure. More technical and auxiliary parts of the text, including the notation are postponed in the Appendix.

\paragraph{Basic notations}
Through the paper we shall adhere to the following notation, see also
\Appx{apx-1}. The position $x$ in the medium is specified through the coordinates
$(x_1,x_2,x_3)$ with respect to a Cartesian reference frame.
The partial derivatives with respect to $x_i$
are denoted by $\partial_i$. We shall also use the microscopic (dilated)
Cartesian reference system of coordinates $(y_1,y_2,y_3)$, therefore the
abbreviations $\pd_i^x = \pd / \pd x_i$ and $\pd_i^y = \pd / \pd y_i$ will be
employed alternatively. 
As usually, the vectors and tensors will be  denoted by  bold letters, for instance,
$\ub(x)$ denotes the velocity vector field depending on the
spatial variable $x$. Moreover, the components of this vector will be denoted
by $u_i$ for $i=1,...,3$, thus $\ub = (u_i)$.  The Einstein summation
convention is used which stipulates implicitly that repeated indices are summed
over. For any two vectors $\ab,\bb$, the inner product is $\ab\cdot\bb$. For
    any  two 2nd order tensors $\Ab,\Bb$ the trace of $\Ab\Bb^T$ is $\Ab:\Bb =
    A_{ij}B_{ij}$.
By $\ol{D}$ we denote the closure of a bounded domain $D$. 
Further, $\nb$ is the unit normal vector defined on a boundary
  $\pd D$, oriented outwards of $D$.
In the context of an interface $\Gamma$ separating domains $\Om_m$ and $\Om_c$, \ie $\Gamma = \ol{\Om_m}\cap\ol{\Om_c}$, normal vector $\nb^{[m]}$ is outward to $\Om_m$ at surface $\Gamma$.
By $\RR$ the real number set is denoted. {Function spaces and other notations are
introduced subsequently in the text and listed in \Appx{apx-1}.} 

\section{Fluid-structure interaction at the microlevel}\label{sec-FSImic}

\subsection{Preliminaries}\label{sec-prelim}

\begin{figure}
\begin{center}
\includegraphics[width=0.6\linewidth]{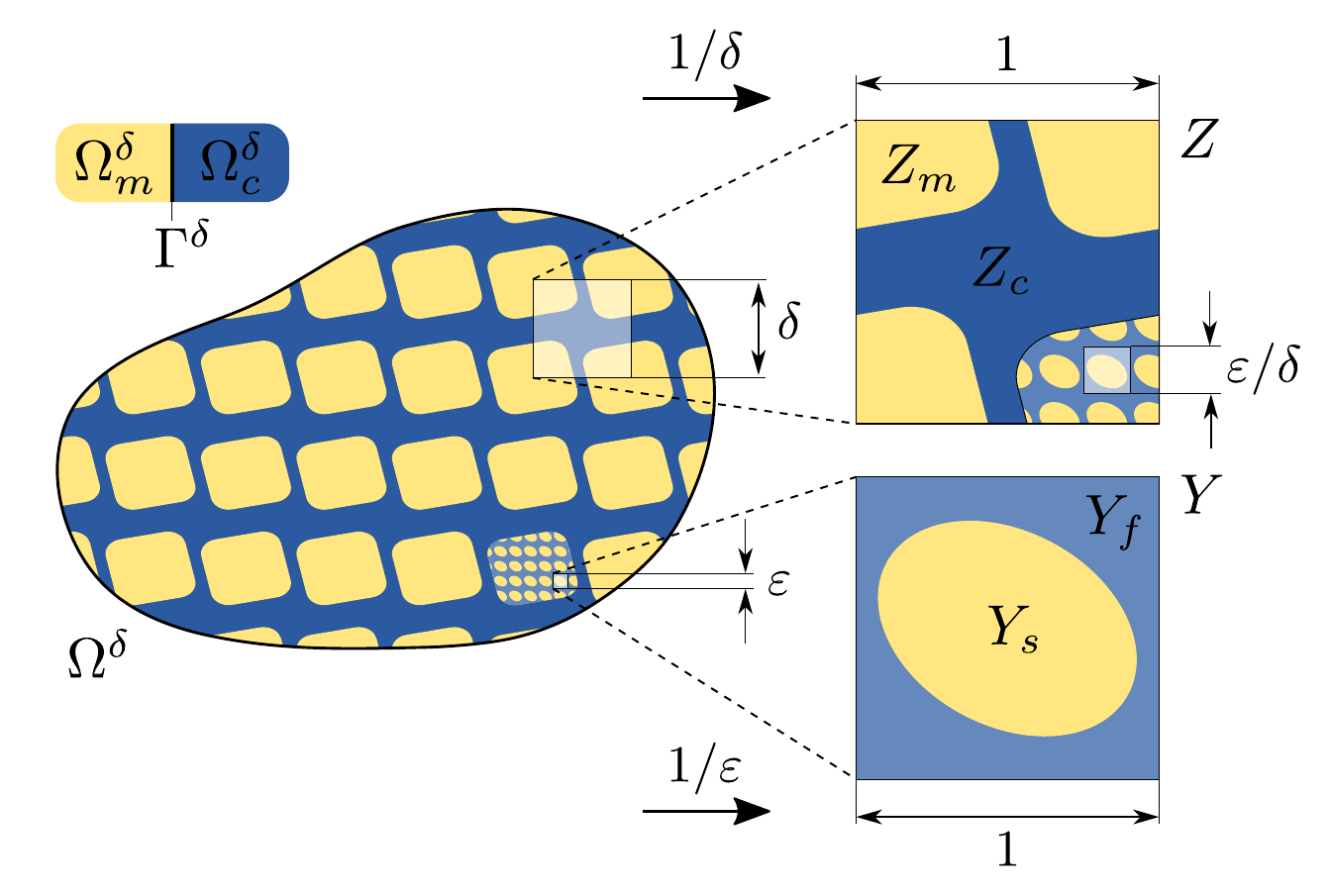}
\caption{Hierarchical porous structure parameterized by $\veps$, the characteristic size of the microporosity, and by $\dlt$ which describes the size of the mesoscopic heterogeneities. Note that $\Om_s^\epsdlt$ and thereby $\Om_m^\dlt$ are connected domains.}
\label{fig1}
\end{center}
\end{figure}

We consider a hierarchically structured porous material consisting of the solid
and fluid phases. The porous material occupies an open bounded domain $\Om \subset
\RR^3$ which can be decomposed in dual way. According to the phases, $\Om$
splits into the fluid $\Om_f$ and solid $\Om_s$ parts, whereby both $\Om_s$ and
$\Om_f$ are connected domains. The hierarchical structure is periodic at the
meso- and micro-scopic scales related to two small parameters $\veps$ and
$\dlt$, see Fig.~\ref{fig1}. At the mesoscopic scale, the periodic structure is
formed by fluid filled channels occupying domain $\Om_c^\dlt$ and by
domain $\Om_m^\dlt = \Om \setminus \Om_c^\dlt$ which is constituted by a microporous
material. In particular, domain $\Om_p^\epsdlt\subset\Om_m^\dlt$ represents
micro pores saturated by fluid, whereas $\Om_s^\epsdlt = \Om_m^\dlt\setminus
\Om_p^\epsdlt$ is the skeleton. To summarize the decompositions,

\begin{equation}\label{eq-mi1}
\begin{split}
\Om & = \Om_m^\dlt \cup \Om_c^\dlt \cup \Gamma^\dlt\;,\\
\mbox{ solid } \quad \Om_s^\epsdlt & \subset \Om_m^\dlt \;,\\
\mbox{ microporosity } \quad \Om_p^\epsdlt & = \Om_m^\dlt \setminus \Om_s^\epsdlt\;,\\
\mbox{ fluid } \quad \Om_f & = \Om_p^\epsdlt \cup \Om_c^\dlt \cup \Gamma_{cp}^\epsdlt\;,\\
\end{split}
\end{equation}
The exterior part $\pd_\ext\Om_s = \pd\Om_s\cap\pd\Om$ of the solid skeleton boundary splits
into two disjoint parts, $\pd_\ext\Om_s = \pd_\sigma \Om_s \cup \pd_u \Om_s$, corresponding to the prescribed type of the boundary conditions. In analogy, we may consider the split of the fluid boundary $\pd_\ext\Om_f = \pd_\sigma \Om_f \cup \pd_v \Om_f$, obviously $\pd_\sigma \Om_f \cap \pd_v \Om_f = \emptyset$.
Further we shall consider $\ol{\Om_s^\epsdlt} \cap \Gamma^\dlt = \emptyset$, 
which means that the mesoscopic ``fictitious'' interface is in the fluid. This property will be used in the 1st level homogenization.

\subsection{Microscopic problem}\label{sec-micpb}
At the microscopic level, we consider the fluid-structure interaction problem for a viscous compressible fluid and elastic  solid with linear material response and linear strains; for the displacement field $\ub(x,t)$ with $x \in \Om_s^\epsdlt$  and time $t \geq 0$, the strain components are $e_{ij}(\ub) = 1/2(\pd_j u_i + \pd_i u_j)$. The elasticity tensor $\Dop = (D_{ijkl})$ satisfies the usual symmetries. 
Following the works \cite{CDG-Stokes-2005}, cf. \cite{RTL-CC2015},  dealing with models of the rigid double porous media,
the viscosity $\eta^\epsdlt$ is given by piece-wise constant function according the micropore size $\veps$:
\begin{equation}\label{eq-mu}
\begin{split}
\eta^\epsdlt = \left \{
\begin{array}{lll}
\veps^2 \bar\eta_p & \mbox{ in } & \Om_p^\epsdlt\;,\\
\eta_c & \mbox{ in } & \Om_c^\dlt\;.
\end{array}
\right.
\end{split}
\end{equation}
This scaling of the viscosity in micropores is the standard consequence of the non-slip boundary condition for the flow velocity on the pore wall, cf. \cite{Hornung1997book}.

In the rest of this section, we suppress the superscripts $^\epsdlt$, or $^\dlt$ to simplify the notation of the domains, parameters, and variables. In principle, however, all material parameters and fields depend on $\veps$ and $\dlt$. 

The problem imposed in $\Om$ at the microlevel is constituted by the following equations and boundary and interface conditions governing the displacement $\ub$ of the solid and both the fluid pressure and velocity fields $(\vb^f,p)$:

\begin{equation}\label{eq-mi2}
\begin{split}
-\nabla\cdot\Dop\eeb{\ub} & = \fb^s\quad \mbox{ in } \Om_s\;,\\
\nb\cdot\Dop\eeb{\ub} & = \nb\cdot\sigmabf^f\quad \mbox{ on }\Gamma_\fsi\;,\\
\nb\cdot\Dop\eeb{\ub} & = \gb^s \quad \mbox{ on } \pd_\sigma\Om_s\;,\\
\ub & = {\bf 0}  \quad \mbox{ on } \pd_u\Om_s\;,\
\end{split}
\end{equation}
\begin{equation}\label{eq-mi3}
\begin{split}
-\nabla\cdot(2\eta\eeb{\vb^f} - p\Ib) & = \fb^f\quad \mbox{ in } \Om_f\;,\\
\nabla\cdot\vb^f & = 0\quad \mbox{ in } \Om_f\;,\\
\vb^f & = \dot\ub\quad \mbox{ on }\Gamma_\fsi\;,\\
\vb^f - \dot{\tilde\ub} =: \wb & = \bar\wb\quad \mbox{ on }\pd_v\Om_f\;, 
\end{split}
\end{equation}
where 
$\sigmabf^f = \veps^2 \bar\eta_p \eeb{\vb^f} - p \Ib$ is the fluid stress, $\fb^{s,f}$ denotes the volume forces in the solid, or in the fluid, and $\gb^s$ is the surface traction stresses acting on the solid part. 

Above, to introduce the relative fluid velocity
$\wb = \vb^f - \dot{\tilde\ub}$ in the fluid-saturated pores $\Om_f^\vepsdel$,
we define a smooth extension $\tilde\ub$ of the displacement field $\ub$ from $\Om_s$ to entire $\Om$, such that 
$\tilde\ub \equiv \ub$ in $\Om_s$.

To introduce the weak formulation of the fluid-structure interaction problem, we use following spaces of admissible and test displacements and relative fluid velocities,
\begin{equation}\label{eq-mi4}
\begin{split}
\Vb_0 & =  \{ \vb \in \Hdb(\Om_s)|\;\vb = {\bf 0}  \mbox{ on }\pd_u\Om_s\}\\
\Wb_0 & = \{ \vb \in \Hdb(\Om_f)|\;\vb = {\bf 0}  \mbox{ on }\pd\Om_f\setminus \pd_\sigma \Om_f\}\;,\quad\quad
\Wb_{\bar\wb}  = \Wb_0 + \bar\wb\;,
\end{split}
\end{equation}
where $\bar\wb$ is a sufficiently regular extension from $\pd_\ext\Om_f$ to $\Om_f$. By $\Hdb(\Om)$ we denote the standard Sobolev space  $\Wb^{1,2}(\Om)$ of vector-valued functions.

The weak formulation of problem \eq{eq-mi2}-\eq{eq-mi3} reads, as follows:
For any time instant $t > 0$, find $(\ub(t,\cdot),\wb(t,\cdot),p(t,\cdot)) \in 
\Vb_0\times\Wb_{\bar\wb} \times H^1(\Om_f))$, such that
\begin{equation}\label{eq-mi5}
\begin{split}
& \int_{\Om_s} \Dop\eeb{\ub}:\eeb{\vb} - \int_\Gamma \nb\cdot\sigmabf^f\cdot\vb
= \int_{\pd_\sigma\Om_s}\gb^s\cdot\vb + \int_{\Om_s}\fb^s\cdot\vb,\quad \vb\in \Vb_0\;,\\
& \int_{\Om_f}2\eta\eeb{\wb+ \dot{\tilde\ub}}:\eeb{\zb} + \int_{\Om_f}\zb\cdot\nabla p  = \int_{\Om_f}
\fb^f\cdot\zb,\quad \zb\in \Wb_0\;,\\
&  \gamma \int_{\Om_f} q \dot p + 
\int_{\Om_f} q \nabla\cdot(\wb+ \dot{\tilde\ub})  = 0\;,\quad \forall q \in L^2(\Om_f)\;.
\end{split}
\end{equation}
Later on, we shall consider more regular test pressure, $q \in H^1(\Om_f)$.
We recall that all functions and parameters involved in  \eq{eq-mi5} depend on the two small parameters $\veps$ and $\dlt$. The fluid domain is decomposed into the microporosity and mesoscopic channels,
$\Om_f^\epsdlt = \Om_p^\epsdlt \cup \Om_c^\dlt \cup \Gamma_{pc}^\epsdlt$.
Passing to the limit with microstructure sized associated with $\veps \rightarrow 0$ give rise to the mesoscopic model reported in Section~\ref{sec-1-homog}. The second level homogenization associated with 
with parameter $\dlt \rightarrow 0$ leads to the macroscopic problem described in Section~\ref{sec-1-homog}.

\subsection{Hierarchical and periodic structure}\label{sec-geom}
The double porosity structure is featured by two length parameters $\ell^\mic$ and $\ell^\meso$ characterizing the size of the micro- and meso-pores, respectively. By virtue of the two-level upscaling, the characteristic lengths $\ell^\mic$ and $\ell^\meso$ are related to  a given macroscopic characteristic length $L$, such that two scale parameters are introduced by $\veps = \ell^\mic/L$ and $\dlt = \ell^\meso/L$.

The porous medium situated in $\Om_m$ is generated as a periodic lattice by repeating the representative volume element (RVE) occupying domain $Y^\veps = \veps Y$. The zoomed cell 
$Y = \Pi_{i=1}^3]0,\bar y_i[ \subset \RR^3$ 
splits into the solid part
occupying domain $Y_s$ and the complementary fluid part $Y_f$, thus
\begin{equation}\label{eq-mi6}
\begin{split}
Y   = Y_s  \cup Y_f  \cup \Gamma_Y \;,\quad
Y_s   = Y \setminus Y_m  \;,\quad
\Gamma_Y   = \ol{Y_s } \cap \ol{Y_f }\;.
\end{split}
\end{equation}
For a given scale $\veps > 0$, $\ell_i = \veps \bar y_i$ is the characteristic size associated with the $i$-th coordinate direction, whereby
also $\veps \approx \ell_i / L$, hence $\ell_i \approx \ell^\mic$ (for all $i=1,2,3$) specifies the microscopic characteristic length $\ell^\mic$. 

At the mesoscopic level, the structure is generated by the periodic cell $\dlt Z$, where $Z =  \Pi_{i=1}^3]0,\bar z_i[$ 
consists of microporous part situated in $Z_m \subset Z$ and of the fluid part $Z_c=Z\setminus Z_m$, \ie $Z_m \cap Z_c = \emptyset$; further, by $\Gamma_Z = \ol{Z_m} \cap \ol{Z_c}$ we denote the interface.
 The size of the channels (mesoscopic pores) is proportional to $\dlt$; in analogy with the microscopic level discussed above, $\dlt \bar z_i \approx \ell^\meso$ (for all $i=1,2,3$).

 Through the paper we designate $\phi_d$ to a volume fraction of the phase $d$ occupying domain $Y_d$, or $Z_d$. In particular,
 $\phi_d = |Y_d|/|Y|$ for $d = s,f$, while $\phi_d= |Z_d|/|Z|$ for $d = m,c$. Obviously, the total volume fraction of the fluid is $\Phi_f = \phi_m\phi_f + \phi_c$, whereas the total  volume fraction of the solid is $\Phi_s = \phi_m\phi_s$, which verify $\Phi_f+\Phi_s = 1$.

\section{Homogenization}\label{sec-homog}
The 1st-level homogenization concerns the asymptotic analysis $\veps\rightarrow 0$ related to the fluid-structure interaction in microporous structure situated in $\Om_m^\dlt$ whereby $\dlt$ is fixed. At the 2nd level homogenization, the asymptotic analysis $\dlt\rightarrow 0$ is related to the mesoscopic heterogeneity and interactions between the fluid in $\Om_c^\dlt$ and the homogenized microporous medium in $\Om_m^\dlt$.

The homogenized model has been derived using the unfolding method of homogenization \cite{Cioranescu-etal-2008}; in \Appx{apx-uf}, the unfolding operator is defined which is employed below to unfold the oscillating functions such that they are expressed in terms of global and local variables (coordinates) describing positions at the upper and lower heterogeneity scales, respectively.  

\subsection{Homogenization -- 1st level}\label{sec-1-homog}
We apply the unfolding operator $\Tuf{~}$, see \Appx{apx-uf}, to transform functions defined in subdomains of  $\Om_m^\dlt$ to functions defined in $\Om_m^\dlt\times Y_d$, $d=s,f$ which is called the unfolded configuration. In the weak formulation \eq{eq-mi5}, we consider test functions supported in $\Om_m^\dlt$ only. Since $\dlt$ is fixed in this 1st level analysis, the superscript $\epsdlt$ will be
replaced by $\veps$ in the notation used through this section. 

The formal  homogenization procedure consists in the following steps:
\begin{list}{}{}
\item (i) Introduce truncated asymptotic expansions which perform as the so-called recovery sequences. The a~priory estimations yield the convergence results for all involved oscillating fields in the unfolded configurations.
\item (ii) The recovery sequences are substituted in the weak formulation transformed in the unfolded configuration. The convergence results yield the limit two-scale problem.
\item (iii) Due to the problem linearity, the two-scale functions can be written using the multiplicative split of characteristic response functions and the mesoscopic fields and their gradients. The characteristic responses which satisfy local problems defined in the micro-configuration $Y$ enable to define the homogenized coefficients involved in the global (mesoscopic) equations of the mesoscopic model.
  \end{list}

\subsubsection{Limit fields and two-scale equations}
The a~priori estimates and the standard results for the homogenization theory \cite{Cioranescu-etal-2008} reveals the following convergences. For any fixed time $t>0$, 
\begin{equation}\label{eq-cnv1}
\begin{split}
\wb^\veps & \cwto \wb_c \quad \mbox{ weakly in  } \Hdb(\Om_c)\;,\\
p^\veps & \cwto p_c \quad \mbox{ weakly in  } L^2(\Om_c) \;,\\
\Tuf{\wb^\veps} & \cwto \wb^0 \quad \mbox{ weakly in  } L^2(\Om_m;\Hdb(Y_f))\;,\\
\Rightarrow\quad \veps \Tuf{\nabla\wb^\veps} & \cwto \nabla_y \wb^0 \quad \mbox{ weakly in  } L^2(\Om_m;\Lb^2(Y^*))\;,\\
\frac{1}{\veps}(\Tuf{p^\veps} - \MeanY{p^\veps}) & \cwto p_m^1 \quad \mbox{ weakly in  } L^2(\Om_m\times Y_f)\;,\\
\MeanYd{f}{p^\veps} & \cwto p_m^0 \quad \mbox{ weakly in  } L^2(\Om_m)\;,\\
\Tuf{\nabla p^\veps} & \cwto \nabla_x p_m^0 + \nabla_y p_m^1 \mbox{ weakly in  } L^2(\Om_m\times Y_f)\;,\\
\Tuf{\ub^\veps} & \cwto \ub^0\quad \mbox{ weakly in  } L^2(\Om_m\times Y_s)\;,\\
\Tuf{\nabla\ub^\veps} & \cwto \nabla_x\ub^0 + \nabla_y\ub^1 \mbox{ weakly in  } L^2(\Om_m\times Y_s)\;,\\
\end{split}
\end{equation}
where $\wb_c \in \Hdb(\Om_c)$, $p_c \in  L^2(\Om_c)$, $\wb^0 \in L^2(\Om_m,\Hpdb(Y_f))$, $p_m^0 \in H^1(\Om_m)$, $p_m^1 \in L^2(\Om_m\times Y_f)$, $\ub^0 \in  \Hdb(\Om_m)$, and $\ub^1 \in  L^2(\Om_m;\Hpdb(Y_s))$ which allows for a smooth extension $\tilde \ub^1 \in  L^2(\Om_m;\Hpdb(Y))$.
All these functions also  depend on time $t$, however, as we shall see, we only treat quasistatic problems where the time dependence is not involved significantly.
Note that the convergence of the gradients $\nabla \ub^\veps$ yields also the convergence of the strains, being the symmetrized gradients. Due to the convergences of $\wb^\veps$ in $ \Hdb(\Om_c)$ and of $\ub^\veps$ in $ \Hdb(\Om_m)$, the Dirichlet boundary conditions are retained, \ie $\ub^0 = 0$ on $\pd_u \Om_m \subset \pd\Om$ and $\wb^0 = 0$ on $\pd_w \Om_c = \pd_\ext \Om_c$.

Moreover, the following properties hold:
\begin{equation}\label{eq-wf1a}
\begin{split}
\nabla_x \cdot\wb_c & = 0 \mbox{  in } \Om_c\;, \\ 
 \nabla_y\cdot \wb^0 & = 0 \mbox{  in } Y_f\;, \quad \wb^0 = 0 \mbox{  on }\Gamma_Y, \\
 \int_{Y_f} p_m^1 & = 0\;. 
\end{split}
\end{equation}
Below more regularity of the pressure fluctuations is needed to compute the gradients, therefore $p_m^1 \in L^2(\Om_m;H_\#(Y_f))$. The ``incompressibility'' condition \eq{eq-wf1a}$_2$
is derived form the limit equation \eq{eq-mi8c}$_1$. Since $\wb^0$ vanish on the interfaces $\Gamma_Y$, we require  $\wb^0 \in L^2(\Om_m,\HOpdb(Y_f))$, where 
\begin{equation}\label{eq-wf1b}
  \HOpdb(Y_f) = \{\wb \in \Hpdbav(Y_f)|\;\wb = 0 \mbox{  on }\Gamma_Y\}\;.
\end{equation}

As the consequence of the convergences \eq{eq-cnv1}, truncated asymptotic expansions of the unfolded unknown fields can be introduced which satisfy the same convergence result. These constitute the recovery sequences in subdomains of $\Om_m^\dlt \equiv \Om_m = \Om_s^\veps\cup\Om_p^\veps$ (note that $\dlt$ is fixed) 
\begin{equation}\label{eq-mi7}
\begin{split}
\Tuf{\chi_s^\veps\ub^\veps} & \approx \ub^0(t,x) + \veps\ub^1(t,x,y)\;,\\
\Tuf{\chi_p^\veps\wb^\veps} & \approx \wb^0(t,x,y)\;,\\
\Tuf{\chi_p^\veps p^\veps} & \approx p_m^0(t,x) + \veps p_m^1(t,x,y)\;,
\end{split}
\end{equation}
where $\chi_d^\veps$ is the characteristic function of subdomain $\Om_d^\veps$, $d = s,p$.
Analogous approximations of the recovery sequences are considered for the test functions $(\vb^\veps, \zb^\veps,q^\veps)$ associated with the unknown functions $(\ub^\veps, \wb^\veps,p^\veps)$. The recovery sequences of the test functions satisfy all convergences in  \eq{eq-cnv1} strongly. These involve limit functions $\vb^0,\vb^1,\zb,q^0$ and $q^1$.

With ansatz \eq{eq-mi7}, in \eq{eq-mi5} we consider only test functions $\vb$ and $q$ which vanish in $\ol{\Om_c}$, thus, being supported in $\Om_m$ only. Upon substituting \eq{eq-mi7} into \eq{eq-mi5}$_1$ and using the fluid stress expression, we get
\begin{equation}\label{eq-mi8a}
\begin{split}
\int_{\Om_m} \intY_{Y_s} \Dop(\eeby{\ub^1} + \eebx{\ub^0}): (\eebx{\vb^0} + \eeby{\vb^1})+  \int_{\Om_m} \intY_{\Gamma_Y}( p_m^1 \vb^0
+ p_m^0 \vb^1)\cdot\nb \\
 -\int_{\Om_m} \vb^0\cdot\intY_{\Gamma_Y}2\bar\eta_p \eeby{\wb^0})\nb^\sx = \int_{\pd_\ext\Om_m}\bar\phi_s \gb\cdot\vb^0  +
\int_{\Om_m}\phi_s\fb^s\cdot\vb^0\;,
\end{split}
\end{equation}
for all $\vb^0 \in \Vb_m$ and $\vb^1 \in C_0^\infty(\Om_m;\Hpdb(Y_s))$. Above, convergence of the interface \lhs integral in \eq{eq-mi5} is explained in Appendix~\ref{apx-2}; the strain $\eeby{\wb^0}$ evaluated on the boundary assumes more regularity on $\wb^0$. However, the sum of the integrals on $\Gamma_Y$ can be replaced by volume forces using the limit of the fluid momentum equation which is obtained from \eq{eq-mi5}$_2$,
\begin{equation}\label{eq-mi8b}
\begin{split}
\int_{\Om_m} \intY_{Y_f} 2\bar{\eta}_p \eeby{\wb^0}:\eeby{\zb} + \int_{\Om_m} \intY_{Y_f} \zb\cdot(\nabla_x p_m^0 + \nabla_y p_m^1) = \int_{\Om_m} \intY_{Y_f}\fb^f\cdot\zb\;,
\end{split}
\end{equation}
for all $\zb \in C_0^\infty(\Om_m;\Hpdb(Y_f))$. Passing into the limit with the mass conservation equation \eq{eq-mi5}$_3$ is a more delicate task. To integrate by parts, we consider test functions $q^0 \in H^1(\Om_m)$ and $q^1 \in C_0^\infty(\Om_m;H_\#^1(Y_m))$ for which the unfolded  mass conservation reads as follows, 
 \begin{equation}\label{eq-mi7c}
\begin{split}
\gamma \int_{\Om_m} \intY_{Y_f} 
(q^0+\veps q^1)( \dot p_m^0+  \veps \dot p_m^1)
- \int_{\Om_m} \intY_{Y_f} (\nabla_x q^0 + \nabla_y q^1)\cdot
(\wb^0 + \dot{\notilde{\ub^0}} + \veps\dot{\wtilde{\ub^1}})\\
+
\int_{\Om_m} \veps^{-1}\intY_{\Gamma_Y} (q^0+\veps q^1)\nb^\fx\cdot( \dot{\notilde{\ub^0}} + \veps\dot{\wtilde{\ub^1}})
+\int_{\pd\Om_m} \chi_f^\veps q^\veps \nb^\mx\cdot \vb^f = 0\;.
\end{split}
 \end{equation}
 Note that here we allow for a nonvanishing $q^0$ on $\Gamma_{cm}$.
 In the limit, upon integrating by parts in the second integral in $Y_f$, the micro- and the meso-scopic problem can be distinguished,
\begin{equation}\label{eq-mi8c}
\begin{split}
\intY_{Y_f} \nabla_y \cdot (\wb^0 + \dot{\notilde{\ub^0}}) q^1 - \intY_{\Gamma_Y}q^1\nb^\fx\cdot(\wb^0 + \dot{\notilde{\ub^0}})
+ \intY_{\Gamma_Y}q^1\nb^\fx\cdot\dot{\notilde{\ub^0}} = 0\quad \mbox{ a.e. in } \Om_m\,,\\
\gamma \int_{\Om_m} \phi_f \dot p^0 q^0 +  \int_{\Om_m} q^0\intY_{\Gamma_Y}\nb^\fx\cdot \dot{\wtilde{\ub^1}}
- \int_{\Om_m} \nabla_xq^0\cdot ( \intY_{Y_f} \wb^0 + \phi_f\dot{\notilde{\ub^0}})
- \int_{\Gamma_{cm}} \phi_f q^0 \vb^f\cdot \nb^\cx \\ = -  \int_{\pd_\ext\Om_m}\bar\phi_f q^0 \bar\vb\cdot \nb^\mx\;.
\end{split}
\end{equation}
As announced above, in \eq{eq-wf1a},  microproblem \eq{eq-mi8c} yields $\nabla_y\cdot \wb^0 =0$ in $Y_f$.

The limit equations \eq{eq-mi5}$_{2,3}$ evaluated for test functions supported in $\Om_c$ are the straightforward consequence of the weak limits of $\wb^\veps$ and $p^\veps$ in domain $\Om_c$, see \eq{eq-me1}.

\subsubsection{Microscopic characteristic responses}
By virtue of the linearity of the local problems arising from \eq{eq-mi8a}, where $\vb^0 = 0$, and \eq{eq-mi8a}, the following split can be introduced
which involves the characteristic responses $\omegabf^{ij},\omegabf^P$, and $\psibf^k,\pi^k$, 
\begin{equation}\label{eq-mi9}
\begin{split}
\ub^1(t,x,y) & = \omegabf^{ij}(y) e_{ij}^x(\ub^0(x)) + \omegabf^P(y) p_m^0(x)\;,\\
\wb^0(t,x,y) & = \psibf^k(y) (\pd_k^x p_m^0(x)-f_k(x))\;,\\
 p_m^1(t,x,y)& = \pi^k(y) (\pd_k^x p_m^0(x)-f_k(x))\;.
\end{split}
\end{equation}
These characteristic responses are obtained by solving local autonomous problems in which we employ the usual elasticity bilinear
form,
$$ 
\aYs{\wb}{\vb} = \intY_{Y_s} (\Dop \eeby{\wb}) : \eeby{\vb}\;.
$$

The  following mutually decoupled problems are to be solved.
\begin{itemize}
\item Find
$\omegabf^{ij}\in \Hpdb(Y_s)$ for any $i,j = 1,2,3$
satisfying
\begin{equation}\label{eq-h5a}
\begin{split}
\aYs{\omegabf^{ij} + \Pibf^{ij}}{\vb} & = 0\;, \quad \forall \vb \in  \Hpdb(Y_s)\;.
\end{split}
\end{equation}
\item Find
$\omegabf^P \in \Hpdb(Y_s)$
satisfying
\begin{equation}\label{eq-h5b}
\begin{split}
\aYs{\omegabf^P}{\vb} & = -\intY_{\Gamma_Y} \vb\cdot \nb^\sx \dSy\;, \quad
\forall \vb \in  \Hpdb(Y_s) \;.
\end{split}
\end{equation}
\item Find $(\psibf^i,\pi^i) \in \Hpdb(Y_f) \times L^2(Y_f)$ for $i = 1,2,3$ such that 
\begin{equation}\label{eq-S3}
\begin{split}
\bar\eta_p\int_{Y_f} \nabla_y \psibf^k: \nabla_y \vb - \int_{Y_f} \pi^k \nabla\cdot \vb & = -\int_{Y_f} v_k\;,\quad \forall \vb \in \Hpdb(Y_f) \;,\\
\int_{Y_f} q \nabla_y \cdot \psibf^k & = 0\;,\quad \forall q \in L^2(Y_f)\;.
\end{split}
\end{equation}
\end{itemize}
Obviously, these problems are classical to establish the poroelastic effective parameters of the Biot continuum. If the microstructure is not perfectly periodic, but it depends on the spatial position, problems \eq{eq-h5a}-\eq{eq-S3} should be solved for almost all  $x\in \Om$, see \eg \cite{rohan-etal-CMAT2015,Rohan-AMC}. It is worth noting that \eq{eq-S3} presents the Stokes problem for incompressible flow, the fluid compressibility $\gamma > 0$ has no influence on the microscopic characteristic response.

\subsubsection{Limit mesoscopic two-scale equations in $\Om_m$ and $\Om_c$}
To establish the mesoscopic model, we use the split \eq{eq-mi9} to express integrals over $Y_d$, $d = s,f$,  involved in the limit equations arising form \eq{eq-mi8a} and  \eq{eq-mi8c} tested by nonvanishing $\vb^0$ and $q^0$, respectively, whereby $\vb^1 = 0$ and $q^1 = 0$. Thus, we get  
\begin{equation}\label{eq-mi10}
\begin{split}
& \int_{\Om_m}\left( \intY_{Y_s} \Dop(\eeby{\ub^1} + \eebx{\ub^0}): \eebx{\vb^0}
+\vb^0\cdot\intY_{\Gamma_Y}( p_m^1\Ib - 2\bar\eta_p \eeby{\wb^0})\nb^\sx\right)  = \\
& \quad\quad\quad \int_{\pd_\ext\Om_m}\bar\phi_s \gb\cdot\vb  +
\int_{\Om_m}\phi_s\fb^s\cdot\vb\;,\\
& \int_{\Om_m} \left( q^0 (\gamma \phi_f\dot p_m^0  + \intY_{\Gamma_Y}\nb^\fx\cdot\dot{\ub}^1)
- \nabla q^0 \cdot(\phi_f \dot{\ub}^0 + \intY_{Y_f} \wb)\right) - \int_{\Gamma_{cm}} q^0 \vb^f\cdot\nb^\cx \phi_f + \int_{\pd_\ext\Om_m} q^0 \bar\phi_f \bar\vb \cdot\nb = 0\;.
\end{split}
\end{equation}
then use:


\begin{equation}\label{eq-mi10a}
\begin{split}
- \int_{\Om_m} \nabla q^0 \cdot \phi_f \dot{\ub}^0 +  \int_{\pd_\ext\Om_m} q^0 \bar\phi_f \bar\vb \cdot\nb
& = \int_{\pd_\ext\Om_m} q^0 (\bar\phi_f \bar\vb - \phi_f \dot{\ub}^0)\cdot\nb
- \int_{\Gamma_{cm}} q^0 \phi_f \dot \ub^0 \cdot \nb^\mx\\
& \quad + \int_{\Om_m} q^0 (\phi_f \nabla \cdot \dot{\ub}^0 + \dot{\ub}^0\cdot
\nabla \phi_f)
\;.
\end{split}
\end{equation}
As the next step, we use \eq{eq-auxmic4} to treat the fluid stress in the interface integral in \eq{eq-mi10}. This leads to term $\vb^0\cdot \phi_f\nabla_x p_m^0$ integrated in $\Om_m$. Integration by parts produces terms $-\phi_f p_m^0 \nabla_x\cdot\vb^0 +\vb^0\cdot p_m^0 \nabla_x\phi_f$; the first term contributes to the coefficient $\Bb$ defined below in \eq{eq-B1a}

\begin{myremark}{rem-per}
  In a case of perfectly periodic media $\nabla_x \phi_f = 0$, however, even in
  this case one must be careful when dealing with surface and volume porosities
  in the case of integration by parts in a volume.
\end{myremark}

Using the characteristic responses \eq{eq-h5a}--\eq{eq-S3} obtained at
the microscopic scale the homogenized coefficients, describing the
effective properties of the deformable porous medium, are given by the
following expressions, \cf \cite{rohan-etal-CMAT2015},

\begin{equation}\label{eq-h8}
\begin{split}
A_{ijkl} = \aYs{\omegabf^{ij} + \Pibf^{ij}}{\omegabf^{kl} + \Pibf^{kl}}\;,\quad
\hat B_{ij} =  -\intY_{Y_s}\dvg_y \omegabf^{ij} = -\aYs{\omegabf^P}{\Pibf^{ij}}\;,\\
M  =  \aYs{\omegabf^P}{\omegabf^P} = \intY_{\Gamma_Y} \omegabf^P\cdot \nb \dSy\;,\quad
K_{ij}  = -\intY_{Y_f} \psi_i^j =  \intY_{Y_f} \nabla_y \psibf^i : \nabla_y \psibf^i \;.
\end{split}
\end{equation}
The coefficients $A_{ijkl}$ and $\hat B_{ij}$ arise from the first integral in \eq{eq-mi10}$_1$, whereas $M$ arises from the interface integral of $\nb^\fx\cdot\dot{\ub}^1$ in \eq{eq-mi10}$_2$ which also produces the alternative expression of $\hat\Bb$. The permeability $\Kb$ is obtained form the average of $\wb^0$ in \eq{eq-mi10}$_2$.
Obviously, the tensors $\Aop = (A_{ijkl} )$, $\hat \Bb = (\hat B_{ij} )$ and
$\Kb = (K_{ij} )$ are symmetric, $\Aop$ adheres all the symmetries of
$\Dop$; moreover, $\Aop$ is positive definite \chE{and $M > 0$}. The
hydraulic permeability $\Kb$ is positive semi-definite in general,
although it is positive definite whenever the channels intersect all
faces of $\pd Y$ and $Y_f$ is connected. Using the volume fraction  $\phi_f = |Y_f|/|Y|$, we define $\Bb$ as follows,
\begin{equation}\label{eq-B1a}
\begin{split}
\Bb & := \hat \Bb + \phi_f \Ib\;.\\
\end{split}
\end{equation}

Upon substituting in \eq{eq-mi10} the expressions involving the two-scale functions using the homogenized coefficients \eq{eq-h8}-\eq{eq-B1a}, we arrive in the weak formulation of the mesoscopic problem. For this we establish  the following functional spaces: 
$\Wb_c^\dlt = \{\wb \in \Hdb(\Om_c^\dlt)|\, \wb = 0 \mbox{ on } \pd_\ext \Om_c^\dlt\}$, $\Vb_m^\dlt =  \{\vb \in \Hdb(\Om_m^\dlt)|\, \vb = 0 \mbox{ on } \pd_u \Om_m^\dlt\}$ and $Q_m^\dlt = H^1(\Om_m^\dlt)$. Also in the notation we drop the superscripts ${}^0$ associated with the mesoscopic limit functions, but replace them by $^\dlt$, since the asymptotic behaviour of the mesoscopic model for $\dlt\rightarrow 0$ is considered in the next section.

\paragraph{Weak formulation of the mesoscopic problem}
Find $(\ub^\dlt,\wb^\dlt, p_m^\dlt, p_c^\dlt)(t,\cdot) \in \Vb_m\times[\Wb_c + \bar\vb] \times H^1(\Om_m^\dlt)\times L^2(\Om_c^\dlt)$  such that
\begin{equation}\label{eq-me1}
\begin{split}
  \int_{\Om_m^\dlt} (\Aop\eeb{\ub^\dlt} - p_m^\dlt \Bb):\eeb{\vb} - \int_{\Gamma_{cm}^\dlt} p_m^\dlt \nb^\cx\cdot\vb
  + \int_{\Om_m^\dlt} p_m^\dlt \vb\cdot \nabla_x \phi_f
  & = \int_{\pd_\ext\Om_m^\dlt}\bar\phi_s \gb\cdot\vb \\
  & \quad + \int_{\Om_m^\dlt}(\phi_s\fb^s + \phi_f \fb^f)\cdot\vb\;,\\
\int_{\Om_m^\dlt}\nabla q_m\cdot \Kb(\nabla p_m^\dlt - \fb^f) + \int_{\Om_m^\dlt}q_m \Bb:\eeb{\dot\ub^\dlt}
+  \int_{\Om_m^\dlt} q \dot\ub^\dlt\cdot \nabla_x \phi_f & \\
+ \int_{\Om_m^\dlt}(M + \gamma\phi_f)\dot p_m^\dlt q_m - \int_{\Gamma_{cm}^\dlt} q_m \wb^\dlt\cdot\nb^\cx
& = \int_{\pd_\ext\Om_m^\dlt} q_m (\phi_f \dot\ub^\dlt-\bar\phi_f\bar\vb) \cdot\nb\;,\\
\int_{\Om_c^\dlt}2\eta_c\eeb{\wb^\dlt+ \dot{\ub}^\dlt}:\eeb{\vthetabf} - \int_{\Om_c^\dlt}p_c^\dlt\nabla \cdot \vthetabf
+ \int_{\Gamma_{cm}^\dlt} p_m^\dlt \nb^\cx\cdot \vthetabf  & = \int_{\Om_c^\dlt}\fb^f\cdot \vthetabf\;,\\ 
\gamma \int_{\Om_c^\dlt} q_c \dot p_c^\dlt + \int_{\Om_c^\dlt} q_c \nabla\cdot(\wb^\dlt+ \dot{\tilde\ub}^\dlt) &  = 0\;, 
\end{split}
\end{equation}
holds for all $\vb\in \Vb_m^\dlt$,  $q_m \in Q_m^\dlt$,  $\vthetabf\in \Wb_c^\dlt$, $q_c \in L^2(\Om_c^\dlt)$.

The \rhs integral in the second equation in \eq{eq-me1} expresses the relative outflow form the matrix part, therefore we introduce 
$\bar w_n^\mic \bar\phi_f := - \nb\cdot(\phi_f \dot\ub-\bar\phi_f\bar\vb)$.



\subsection{Homogenization -- 2nd level}\label{sec-2-homog}
We pursue the analogous procedure as the one applied at the 1st level upscaling. In this section, the unfolding operator $\Tufd{~}$ is employed. In the context of decomposition \eq{eq:3a}, we shall consider $\{x/\dlt\}_Z = z \in Z$.
To derive 
the limit model describing the medium behaviour at the macroscopic scale, asymptotic analysis of \eq{eq-me1} for $\dlt\rightarrow 0$ is carried out which yields
the local mesoscopic problems for two-scale functions and the limit macroscopic equations. w. Note that all functions involved in \eq{eq-me1} depend on $\dlt$ as the result of the scale-dependent partitioning of $\Om$ into its subparts $\Om_m^\dlt$ and $\Om_c^\dlt$.

\subsubsection{Mesoscopic heterogeneous structure}
The mesoscopic cell is decomposed into the ``microporous'' matrix and the mesoscopic channels, $Z = Z_m \cup Z_c \cup \Gamma_Z$, which are separated by the interface $\Gamma_Z$.
The global domain $\Om$ generated by $\dlt Z$ as a periodic lattice is decomposed into the corresponding parts, $\Om = \Om_m^\dlt \cup \Om_c^\dlt \cup \Gamma_{cm}^\dlt$. 
We recall that the interface $\Gamma_{cm}^\dlt$ is ``immersed'' in the fluid; this is is the assumption involved in the 1st level homogenization. Therefore, also the generating interface $\Gamma_Z^\dlt$ which is the $d-1$ dimensional manifold in the real-sized cell part $\dlt Z = Z^\dlt$ is associated to the fluid.  

From the mesoscopic problem \eq{eq-me1}, the following convergences of unfolded functions can be obtained;
for the sake of simplicity of the notation, we designate the limit functions of displacements by the same symbols $\ub^0$ and $\ub^1$, as in the 1st level upscaling. Also the results involving functions of time are presented for any fixed time $t>0$.
\begin{equation}\label{eq-cnv2}
  \begin{split}
    \Tufd{\ub^\dlt} & \cwto \ub^0\quad \mbox{ weakly in  } L^2(\Om \times Z_m)\;,\\
    \Tufd{\nabla\ub^\dlt} & \cwto \nabla_x\ub^0 + \nabla_z\ub^1 \mbox{ weakly in  } L^2(\Om\times Z_m)\;,\\
    \Tufd{\wb^\dlt} & \cwto \wb^0\quad \mbox{ weakly in  } L^2(\Om \times Z_c)\;,\\
    \Tufd{\nabla\wb^\dlt} & \cwto \nabla_x\wb^0 + \nabla_z\wb^1 \mbox{ weakly in  } L^2(\Om\times Z_c)\;,\\
 \Tufd{p_m^\dlt} & \cwto p^0   \mbox{ weakly in  } L^2(\Om\times Z_m)\;,\\
 \Tufd{\nabla p_m^\dlt} & \cwto \nabla_x p^0 + \nabla_z p^1 \mbox{ weakly in  } L^2(\Om\times Z_m)\;,\\
 \Tufd{p_c^\dlt} & \cwto \hat p  \mbox{ weakly in  } L^2(\Om\times Z_c)\;,\\
 {\dlt}\Tufd{p_c^\dlt} & \cwto \nabla_z \hat p  \mbox{ weakly in  } L^2(\Om\times Z_c)\;,
\end{split}
\end{equation}
where $\ub^0,\wb^0  \in  \Hdb(\Om)$, $\ub^1 \in  L^2(\Om;\Hpdb(Z_m))$,  $\wb^1 \in  L^2(\Om;\Hpdb(Z_c))$, $p^0 \in H^1(\Om)$,  $p^1 \in L^2(\Om; H_\#^1(Z_m))$, $\hat p \in L^2(\Om; H_\#^1(Z_c))$. As in the 1st level upscaling, we rely on existence of smooth extensions of $\ub^1(x,\cdot)$ from $Z_m$ to entire $Z$; this is needed to introduce the fluid velocity in $Z_c$ in terms of the relative velocity $\wb^1$. 

 Formal asymptotic truncated expansions of the unfolded unknown fields which
satisfy the same convergence results can be introduced. 
The following approximations constitute the recovery sequences which are then substituted in the unfolded equations \eq{eq-me1},
\begin{equation}\label{eq-me2}
\begin{split}
\Tufd{\chi_m^\dlt\ub^\dlt} & \approx \ub^0(t,x) + \dlt\ub^1(t,x,z)\;,\\
\Tufd{\chi_c^\dlt\wb^\dlt} & \approx \wb^0(t,x) + \dlt\wb^1(t,x,z)\;,\\
\Tufd{\chi_m^\dlt p_m^\dlt} & \approx p^0(t,x) + \dlt p^1(t,x,z)\;,\\
\Tufd{\chi_c^\dlt p_c^\dlt} & \approx \hat p(t,x,z)\;.
\end{split}
\end{equation}
Analogous expressions of the test functions $(\vb^\dlt, \vthetabf^\dlt,q_m^\dlt,q_c^\dlt)$ associated with unknown functions  $(\ub^\dlt,\wb^\dlt,p_m^\dlt,p_c^\dlt)$ are considered.

We proceed by deriving the limit two-scale equations from \eq{eq-me1}. For this, \eq{eq-me2} and expressions of the test functions are substituted in \eq{eq-me1}. Passing to the limit $\dlt\rightarrow 0$ using the convergences state above, the local  mesoscopic problems and global macroscopic problems can be distinguished by choosing suitable combinations of the limit test functions. In contrast to the convergence of the volume integrals involved in  \eq{eq-me1}, the limit expressions of all the interface boundary integrals need more attention, see the details given in Appendix~\ref{apx-3}.

\begin{myremark}{rem-bc}
  In these expression, we obtain traces of the pressure $\bar p^0$ and  velocity $\bar\wb$ on $\pd\Om$. In the weak formulation, theses traces are prescribed on subparts $\pd_p\Om$ and $\pd_w\Om$, respectively, whereas they are not needed on the complementary parts of the boundary due to the Dirichlet boundary conditions. They are adhered due to the weak convergences of $\ub^\dlt$ and $\wb^\dlt$ in $ \Hdb(\Om)$, hence  $\ub^0 = 0$ on $\pd_u \Om \subset \pd\Om$ and $\wb^0 = 0$ on $\pd_w \Om$. Therefore, we introduce the following spaces: $\Vb_0^H = \{\vb \in \Hdb(\Om)|\, \vb = 0 \mbox{ on } \pd_u \Om\}$, $\Wb_0^H = \{\wb \in \Hdb(\Om)|\, \wb = 0 \mbox{ on } \pd_w \Om\}$, and $Q_0^H = \{q \in H^1(\Om)|\, q = 0 \mbox{ on } \pd_p \Om\}$. Note that for simplicity, in \eq{eq-mi3}, we considered $\bar \wb$ prescribed on entire $\pd_\ext \Om_f^\epsdlt$, which leads to  $\pd_w \Om = \pd\Om$ and $\pd_p\Om = \emptyset$. However, formally we can consider a more general combinations of the boundary conditions when passing to the limit.
\end{myremark}

We now present the limit two-scale equations, where $\vb^0(x)$ and $\vb^1(x,z)$ are test displacements, $\vthetabf^0(x)$ and $\vthetabf^1(x,z)$ are test seepage velocities, and $q^0(x), q^1(x,z)$ and $\hat q(x,z)$ are test pressures.

\paragraph{Local mesoscopic problems}
These equations are obtained for almost all $x \in \Om$ with vanishing macroscopic test functions, \ie $\vb^0, \vthetabf^0 = 0$ and $q^0 = 0$ in $\Om$,
\begin{equation}\label{eq-me10}
\begin{split}
\intY_{Z_m} \Aop(\eebz{\ub^1} +\eebx{\ub^0} : \eebz{\vb^1} - \intY_{Z_m}p^0 \Bb: \eebz{\vb^1} 
+ \intY_{\Gamma_Z} p^0 \nb^\mx\cdot\vb^1
& = 0\;, \\
\intY_{Z_m} \nabla_z q^1 \cdot \Kb(\nabla_z p^1 + \nabla_x p^0 - \fb^f) - \intY_{\Gamma_Z} q^1 \nb^\cx  \cdot\wb^0 & = 0\;, \\
\intY_{Z_c}2\eta_c \left[\eebz{\dot{\tilde\ub}^1 + \wb^1} + \eebx{\dot{\tilde\ub}^0 + \wb^0}
\right]:\eebz{\vthetabf^1} - \intY_{Z_c} \hat p \nabla_z\cdot\vthetabf^1 + \intY_{\Gamma_Z}\vthetabf^1 \cdot\nb^\cx p^0 & = 0 \;, \\
\intY_{Z_c}\left(\nabla_z \cdot (\wb^1 + \dot{\tilde\ub}^1) + 
\nabla_x\cdot(\dot{\tilde\ub}^0 +\wb^0) \right)\hat q + \gamma\intY_{Z_c}\dot{\hat p} \hat q & = 0  \;, \\
\end{split}
\end{equation}
for all test functions $\vb^1\in \Hpdb(Z_m)$, $q^1 \in H_\#^1(Z_m)$, $\wb^1 \in \Hpdb(Z_c)$, and $\hat q \in L^2(Z_c)$.

\subsection{Macroscopic problem}\label{sec-mac}
These equations are obtained with vanishing two-scale test functions, \ie  $\vb^1, \vthetabf^1 = 0$, $\hat q = 0$ and $q^1 = 0$ in $\Om \times Z$,
\begin{equation}\label{eq-ma1-i}
\begin{split}
&\int_\Om \intY_{Z_m} \left(
\Aop(\eebx{\ub^0} + \eebz{\ub^1}):\eebz{\vb^0}
- p^0 \Bb:\eebx{\vb^0}
\right) 
+ \int_\Om \vb^0\cdot\intY_{Z_m}\nabla_z p^1\\
& -\int_{\Om} \phi_c \nabla_x \cdot(p^0 \vb^0)  
= \int_\Om \phi_m (\phi_s\fb^s + \phi_f\fb^f)\cdot\vb^0 + \int_{\pd\Om}(\bar\phi_m \bar\phi_s \gb - \phi_c\nb p^0) \cdot\vb^0 \;,\\
&\int_\Om \intY_{Z_m} q^0 \Bb:(\eebx{\dot\ub^0} + \eebz{\dot\ub^1}) 
+ \int_\Om \intY_{Z_m} \nabla q^0 \cdot \Kb (\nabla_z p^1 + \nabla_x p^0 -\fb^f)
+ \int_\Om \intY_{Z_m} q^0 M  \dot p^0  \\
& 
- \int_\Om q^0 \intY_{\Gamma_Z} \nb^\cx\cdot  \wb^1 - \int_\Om \phi_c\nabla_x\cdot ( q^0\wb^0) =
-\int_{\pd\Om} q^0 (\bar\phi_m\bar\phi_f \bar w_n^\mic + \bar\phi_c \bar w_n^\meso) \;,\\
&\int_\Om \intY_{Z_c} 2\eta_c \left(\eebx{\wb^0+\dot{{\ub}}^0}
+ \eebz{\wb^1 + \dot{\tilde{\ub}}^1}
\right):\eebx{\vthetabf^0}  - \int_\Om \intY_{Z_c} \hat p \nabla_x\cdot\vthetabf^0
+ \int_\Om \vthetabf^0\cdot\intY_{\Gamma_Z}p^1 \nb^\cx \\ 
& +\int_\Om \phi_c \nabla_x \cdot ({p}^0 \vthetabf^0)   = 
\int_\Om \phi_c\fb^f\cdot\vthetabf^0 + \int_{\pd\Om} \bar\phi_c {p}^0 \nb\cdot\vthetabf^0 
\;,
\end{split}
\end{equation}
for all $(\vb^0,q^0,\vthetabf^0) \in \Hdb(\Om)\times H_\#^1(\Om) \times \Hdb(\Om)$, where  $\bar w_n^\mic$ and $\bar w_n^\meso$ are the relative fluid  velocities associated with the micro- and mesoporosities; these can be given according to the boundary conditions.

The further procedure in deriving the homogenized problem is to introduce characteristic responses defined in the representative unit cell $Z$. Because of the time dependence of the two scale functions $\ub^1(x,y,t),\wb^1(x,y,t),p^1(x,y,t)$ and $\hat p(x,y,t)$ are coupled in time with the macroscopic functions $\ub^0(x,t),\wb^0(x,t)$ and $p^0(x,t)$. Therefore, the split must be introduced for the functions transformed using the Laplace transformation \wrt the time. The macroscopic model then involves time convolutions with kernels constituted by homogenized coefficients defined in terms of the characteristic responses, solutions of the evolutionary local problems. In this paper we restrict to a simpler situation: when the fluid is incompressible, the ``scale-decoupling'' procedure can be done without need to use the Laplace transformation.

\section{Macroscopic model for incompressible fluids, $\gamma = 0$}\label{sec-mac-i}
In the rest of the paper we shall consider the mesoscopic medium governed by model \eq{eq-me1} with  $\gamma = 0$ which describes incompressible fluids saturating the micro- and meso-scopic pores. This restriction leads to a less complicated procedure of introducing the effective material properties. It will be shown that the macroscopic coefficients are time independent and the evolutionary macroscopic problem involves time derivative only rather than integro-differential operators which feature the macroscopic problem in a general case. We shall consider such a more general case in our further work.

\subsection{Local characteristic responses}\label{sec-meso-loc}
We consider local two-scale problem \eq{eq-me10}, where $\gamma = 0$.
Due to the vanishing term with $\dot{\hat p} $, it is possible to introduce the characteristic responses using the standard split in the time domain,                  
\begin{equation}\label{eq-me11-i}
\begin{split}
\ub^1 & =  \omegabf^{ij} e_{ij}^x(\ub^0) +  \omegabf^P p^0 \;,\\
p^1 & =  \pi^k(\pd_k^x p^0 - f_k^f) +  \vphi^k w_k^0\;,\\
\wb^1 & =  \psibf^{ij} e_{ij}^x(\wb^0) +  \chibf^{ij} e_{ij}^x( {\dot\ub}^0)+  \psibf^P p^0 + \xibf^P \dot p^0\;,\\
\hat p & =  \hat \vphi^{ij}  e_{ij}^x(\wb^0) +  \hat\eta^{ij} e_{ij}^x( {\dot\ub}^0)+  \hat\eta^P p^0+  \hat\zeta^P \dot p^0 \;,\\
\end{split}
\end{equation}
where the all the characteristic responses $\omegabf^{ij}, \omegabf^P, \pi^k,  \chibf^{ij}, \psibf^P,  \hat \vphi^{ij}, \hat\eta^{ij}, \hat\eta^P$ and $\hat\zeta^P$ are $Z$-periodic functions defined in $Z_m$ or $Z_c$. All these functions are independent of time.

The local mesoscopic characteristic problems are defined in terms of bilinear forms:
\begin{equation}\label{eq-me12}
\begin{split}
\aYm{\ub}{\vb} & = \intY_{Z_m} \Aop \eebz{\ub}:\eebz{\vb} \;, \\
\bYm{p}{\vb} & = \intY_{Z_m} p \Bb:\eebz{\vb} \;, \\
\cYm{p}{q} & = \intY_{Z_m} \nabla_z q\cdot \Kb\nabla_z p\;,\\
\ipZc{u}{w} & = \intY_{Z_c} u w\;,\\
\ipZc{\eebz{\ub}}{\eebz{\wb}} & = \intY_{Z_c}\eebz{\ub}:\eebz{\wb}\;.
\end{split}
\end{equation}

\paragraph{Characteristic responses in the matrix part $Z_m$}
Find $\omegabf^{ij},\omegabf^P\in\Hpdb(Z_m)/\RR^3$, and $\pi^k, \vphi^k \in H_\#^1(Z_m)$ such that
\begin{equation}\label{eq-me13a}
\begin{split}
\aYm{\omegabf^{ij}}{\vb} & = - \aYm{\Pibf^{ij}}{\vb}\quad \forall \vb\in\Hpdb(Z_m)\;,\\
\aYm{\omegabf^P}{\vb} & = \bYm{1}{\vb}-\intY_{\Gamma_Z} \nb^\mx\cdot\vb\quad \forall \vb\in\Hpdb(Z_m)\;,\\
\cYm{\pi^k}{q} & = - \cYm{z_k}{q}\quad \forall q \in H_\#^1(Z_m)\;,\\
\cYm{\vphi^k}{q} & = {\intY_{\Gamma_Z} q n_k^\cx} \quad \forall q \in H_\#^1(Z_m)\;.
\end{split}
\end{equation}
Obviously \eq{eq-me13a} comprises four mutually decoupled problems.

\paragraph{Characteristic responses in the channels $Z_c$}
Although \eq{eq-me11-i} introduces corrector functions to four macroscopic fields, flow strains $e_{ij}^x(\wb^0)$, displacement strains $e_{ij}^x(\ub^0)$, pressure $p^0$ and its rate $\dot p^0$, it will be shown that only one problem for characteristic response associated with $e_{ij}^x(\wb^0)$ must be solved.

Find $(\psibf^{ij},\hat\vphi^{ij}) \in \Hpdb(Z_c)/\RR^3\times L^2(Z_c)$, such that,
\begin{equation}\label{eq-me15b-i}
\begin{split}
2\eta_c \ipZc{\eebz{\psibf^{ij}}}{\eebz{\vthetabf}} - \ipZc{\hat\vphi^{ij}}{\nabla_z\cdot\vthetabf}
& = -2\eta_c \ipZc{\eebz{\Pibf^{ij}}}{\eebz{\vthetabf}}
\;,\\
\ipZc{\nabla_z\cdot\psibf^{ij}}{q}  & = 
- \ipZc{\nabla_z\cdot\Pibf^{ij}}{q} \;,
\end{split}
\end{equation}
for all $(\vthetabf,q) \in \Hpdb(Z_c)\times L^2(Z_c)$.

The following proposition states that other characteristic responses involved in \eq{eq-me11-i} have either trivial solutions, or can be expressed using the solutions of \eq{eq-me15b-i}.

\begin{myproposition}{prop-1}
 Let $\tilde\omegabf^{ij}$ and $\tilde\omegabf^P$ be any smooth $Z$-periodic extensions of responses $\omegabf^{ij}$ and $\omegabf^P$ defined in \eq{eq-me13a}$_1$ and  \eq{eq-me13a}$_2$ , respectively. Further denote $\tilde\psibf^{ij} := \chibf^{ij} + \tilde\omegabf^{ij}$ and $\tilde \vphi^{ij} := \hat \vphi^{ij}$.

\begin{list}{}{}
\item (i)\quad Characteristic responses $\psibf^{ij} \equiv \tilde\psibf^{ij}$ and $\hat\eta^{ij}\equiv \tilde \vphi^{ij}$ are solutions of \eq{eq-me15b-i}.
\item (ii)\quad   
   The couple $(\tilde\psibf^{ij},\hat \eta^{ij})$ satisfies problem \eq{eq-me15b-i}, thus $\tilde\psibf^{ij} \equiv \psibf^{ij}$ and $\hat\eta^{ij}\equiv \tilde \vphi^{ij}\equiv\hat \vphi^{ij}$.

\item (iii) \quad $\psibf^P \equiv 0$ and $\hat\eta^P \equiv  1$.
\item (iv)\quad $\xibf^P + \tilde\omegabf^P \equiv  0$ and also $\hat \zeta^P \equiv  0$.
    
\end{list}

\end{myproposition}

\begin{myproof}{Proposition \ref{prop-1}.}
  The proof is an easy consequence of the autonomous problems
  \eq{eq-me15b-i} and \eq{eq-me15a-i}-\eq{eq-me15d-i} which are introduced by virtue of the split \eq{eq-me11-i} substituted in the local problem \eq{eq-me1}. This yields (i). Further note that $\tilde\psibf^{ij}\in \Hpdb(Z_c)$.

  Then (ii) follows upon introducing $\tilde\psibf^{ij}$ and $\tilde \vphi^{ij}$ in \eq{eq-me15a-i} which, thus, become identical with  \eq{eq-me15b-i}.

  To show (iii), the \rhs in \eq{eq-me15c-i} can be rewritten in terms of $\nabla_z \thetabf$. Then putting $\vthetabf := \psibf^P$ and using  \eq{eq-me15c-i}$_2$ yields $\psibf^P\equiv 0$ due to coercivity of the bilinear form $\ipZc{\cdot}{\cdot}$ in $\Hpdb(Z_c)/\RR^3$. Hence $\hat\eta^P \equiv  1$.
  
 To show (iv), use $\vthetabf := \xibf^P + \tilde\omegabf^P$ and \eq{eq-me15d-i}$_2$. The rest follows by similar arguments as those employed in the proof of (iii).

\end{myproof}  

\subsection{Homogenized coefficients and the macroscopic model}\label{sec-meso-HC}
The homogenized coefficients are obtained from \eq{eq-ma1-i} which is not affected by the fluid incompressibility -- this phenomenon influences only equations of the local mesoscopic problem and, thereby, also the effective properties. All the homogenized coefficients (HC) are time-independent. Below we present only lists of expressions which are derived in Appendix~\ref{apx-4}. As usually, coupling coefficients appear in the upscaled system of equations which satisfy the reciprocity principles; the symmetry and reciprocity properties are introduced in Proposition~\ref{prop-2}.

The first set of HC is identified in \eq{eq-ma1-i}$_1$ upon substituting there the split \eq{eq-me11-i}.
\begin{equation}\label{eq-ma2-i}
\begin{split}
  \Acal_{ijkl} & = \intY_{Z_m} \Aop \eebz{\Pibf^{kl} + \omegabf^{kl}}:\eebz{\Pibf^{ij}}\\
  & = \intY_{Z_m} \Aop \eebz{\Pibf^{kl} + \omegabf^{kl}}:\eebz{\Pibf^{ij} +  \omegabf^{ij}}\;,\\
\Bcal_{ij}^* & = \phi_c \delta_{ij} + \phi_m \Bb_{ij} - \intY_{Z_m} \Aop \eebz{\omegabf^P}:\eebz{\Pibf^{ij}}\\
           &    = \phi_c \delta_{ij} + \bYm{1}{\Pibf^{ij} +  \omegabf^{ij}} - \intY_{\Gamma}\nb^\mx\cdot\omegabf^{ij}\;,\\
\Hcal_{ij}^* & = \intY_{\Gamma}\vphi^i n_j^\mx = -\cYm{\vphi^i}{\vphi^j} \;,\\ 
\Qcal_{ij}^* & = \intY_{\Gamma}\pi^j n_i^\mx \;.
\end{split}
\end{equation}
The alternative expressions are due to the local problems \eq{eq-me13a}. In the same way, the second set of HC is identified in \eq{eq-ma1-i}$_2$. 
\begin{equation}\label{eq-ma3-i}
\begin{split}
  \Bcal_{ij} & = \phi_c \delta_{ij} + \bYm{1}{\Pibf^{ij} +  \omegabf^{ij}} - \intY_{\Gamma}\nb^\mx\cdot\omegabf^{ij}\;,\\
\Kcal_{ij} & = \intY_{Z_m} \nabla_z(z_j + \pi^j)\cdot\Kb\nabla_zz_i = 
\intY_{Z_m} \nabla_z(z_j + \pi^j)\cdot\Kb\nabla_z(z_i + \pi^i)\;,\\
\Qcal_{ij} & = \intY_{Z_m}  \nabla_z z_i \cdot \Kb\nabla_z\vphi^j = 
-\intY_{Z_m}  \nabla_z \pi^i \cdot \Kb\nabla_z\vphi^j = \intY_{\Gamma_Z} \pi^i n_j^\mx\;,\\
\Mcal & = \intY_{Z_m}   M  + \bYm{1}{\omegabf^P} + \intY_{\Gamma} \nb^\cx\cdot\omegabf^P\;.
\end{split}
\end{equation}
Finally the third group of HC yields from  \eq{eq-ma1-i}$_3$.
\begin{equation}\label{eq-ma4-i}
\begin{split}
  \Scal_{ijkl} & = 2\eta_c\intY_{Z_c}  \eebz{\psibf^{kl} + \Pibf^{kl}}:\eebz{\Pibf^{ij}} - \intY_{Z_c}{\hat\vphi}^{kl}\nabla_z\cdot \Pibf^{ij}\\
  & = 2\eta_c\ipZc{\eebz{\Pibf^{ij} + \psibf^{ij}}}{\eebz{\Pibf^{kl} + \psibf^{kl}}}\;,\\
\Hcal_{ij} & = \intY_{Z_c}\vphi^j n_i^\cx = \cYm{\vphi^i}{\vphi^j}\;,\\
\Qcal_{ij}^* & = - \intY_{\Gamma_Z}\pi^j n_i^\cx =  \intY_{\Gamma_Z}\pi^j n_i^\mx\;.
\end{split}
\end{equation}

All coefficients are symmetric \wrt indices related to strain and strain rate tensors, \ie $a_{ij} = a_{ji}$; this is an obvious consequence of $e_{kl}^z{\Pi^{ij}} = 1/2 (\pd_l^z z_j\dlt_{ik} + \pd_k^z z_j\dlt_{il}) = 1/2(\dlt_{jl}\dlt_{ik} + \dlt_{kj}\dlt_{il}) = e_{kl}^z{\Pi^{ji}}$. Furthermore, reciprocity relationships hold. 

\begin{myproposition}{prop-2}
 \begin{list}{}{}
\item (i)\quad The following equalities hold:
\begin{eqnarray}
  \Scal_{ijkl} & = &\Scal_{klij} =  \Scal_{lkij}\;,\label{eq-ma5-1}\\
  \Acal_{ijkl} & = &\Acal_{klij} =  \Acal_{lkij}\;,\label{eq-ma5-1a}\\
\Hcal_{ij} & = &- \Hcal_{ij}^* =  \Hcal_{ji}\;,\label{eq-ma5-2}\\
\Qcal_{ij} & = & \Qcal_{ji}^*\;,\label{eq-ma5-3}\\
\Bcal_{ij} & = & \Bcal_{ij}^*\;.\label{eq-ma5-4}
 \end{eqnarray}
 Moreover, $(\Acalbf\ab):\ab > 0$ and $(\Scalbf\ab):\ab > 0$ for any second-order tensor $\ab = (a_{ij})$, and $\Mcal > 0$. 
\item (ii)\quad If an isotropic (and homogeneous) permeability is obtained at the mesoscopic scale, $K_{ij} = \kappa \delta_{ij}$, then $\Qcal_{ij} = \Qcal_{ji}$ is symmetric and $\Qcalbf = \Qcalbf^*$. Moreover $\Qcal_{ij}=\kappa \Hcal_{ij}$.
 \end{list}  
\end{myproposition}

It is worth noting that the symmetry of $\Qcalbf$ depends on geometry of the microchannels $Y_f$, whereas it is independent of the geometry of the mesoscopic channels geometry.

\begin{myproof}{Proposition \ref{prop-1}}
  (i) To prove \eq{eq-ma5-1}, thereby also the positive definiteness of $\Scalbf$, substitute $\vthetabf = \hat\psibf^{kl}$ in \eq{eq-me15b-i}$_1$ and consider \eq{eq-me15b-i}$_2$ with indices $kl$ and with substituted $q = \hat \psi^{ij}$. Upon summation of the two equations, the identity is added to the primary expression for $\Scal_{ijkl}$ in \eq{eq-ma4-i}$_1$. For proving the other symmetries \eq{eq-ma5-1a}-\eq{eq-ma5-4}, positive definiteness of $\Acalbf$, and positivity of $\Mcal$,  identities arising from \eq{eq-me13a} with choosing suitable test functions $\vb$ and $q$ are used.
  (ii) From \eq{eq-ma3-i}, using the assumptions of the isotropy and homogeneity, we get 
  \begin{equation*}
\begin{split}
\Qcal_{ij} = \cYm{z_i}{\vphi^j} = \kappa \intY_{Z_c} \pd_i^z \vphi^j = \kappa \intY_{\Gamma_Z} \vphi^j n_i^\cx = \kappa  \cYm{\vphi^i}{\vphi^j} = \kappa \Hcal_{ij}\;.
\end{split}
  \end{equation*}
\end{myproof}

Further we define 
\begin{equation}\label{eq-ma6-i}
\begin{split}
  \Pcal_{ij} & = \phi_c\delta_{ij} - \Qcal_{ij} \;,\\
  \Pcal_{Jo}^* & = \phi_c\delta_{ij} - \Qcal_{ji}^* \;,
\end{split}
\end{equation}
hence $\Pcal_{ij}  = \Pcal_{ji}^*$. To employ $\Pcalbf^*$ in the equilibrium equation of \eq{eq-ma10-i}, we use 
\begin{equation}\label{eq-ma7-i}
\begin{split}
-\int_\Om \vb^0\cdot(\Pcalbf^*\nabla_x p^0 + \Qcalbf^*\fb^f) = -\int_\Om \vb^0\cdot\Pcalbf^*(\nabla_x p^0 -\fb^f) - \int_\Om \phi_c\fb^f\cdot \vb^0\;.
\end{split}
\end{equation}

\paragraph{Macroscopic model} Upon substituting in equations \eq{eq-ma1-i} the integrals in $Z_m$ and $Z_c$ using the homogenized coefficients, the weak formulation of the macroscopic problem is obtained. Find  $(\ub^0,p^0,\wb^0) \in \Vb_0^H\times [Q_0 + \bar p^0]\times [\Wb_0^H +\bar\wb]$,
such that the following equations hold 
\begin{equation}\label{eq-ma10-i}
\begin{split}
\int_\Om \left(\Acalbf\eebx{\ub^0} - p^0\Bcalbf^*\right):\eebx{\vb^0} 
- \int_\Om \vb^0\cdot\Pcalbf^*(\nabla_x p^0- \fb^f) & \\+ \int_\Om \vb^0\cdot\Hcalbf^*\wb^0 
+ \int_{\pd_\sigma\Om} \bar\phi_c \bar p^0\nb\cdot\vb^0
& = 
\int_\Om \fb^\blk \cdot\vb^0 + \int_{\pd_\sigma\Om}\bar\phi_m \bar\phi_s \gb\cdot\vb^0\;,\\
\int_\Om q^0 \Bcalbf:\eebx{\dot\ub^0} + \int_\Om \nabla_x q^0 \cdot\left(\Kcalbf(\nabla_x p^0 - \fb^f) 
- \Pcalbf\wb^0\right)
+\int_\Om q^0  \Mcal\dot p^0  & = -\int_{\pd_w\Om} q^0 (\bar\phi_m\bar\phi_f \bar w_n^\mic + \bar\phi_c \bar w_n^\meso)\;,\\
\int_\Om\eebx{\thetabf^0}:\Scalbf\eebx{\wb^0+\dot\ub^0} + \int_\Om \thetabf^0\cdot\Pcalbf(\nabla_x p^0 - \fb^f)
+ \int_\Om \thetabf^0\Hcalbf\wb^0 & = \int_{\pd_p\Om} \bar\phi_c \bar p^0\nb\cdot\thetabf^0\;,
\end{split}
\end{equation}
for all $(\vb^0,q^0,\thetabf^0) \in \Vb_0^H \times Q_0^H \times  \Wb_0^H$, where
$\fb^\blk =  \phi_m(\phi_s\fb^s + \phi_f\fb^f) +  \phi_c\fb^f$ is the bulk volume force acting on the fluid-solid mixture.
The spaces involved in this formulation were introduced in Remark~\ref{rem-bc}. For clarity, tensors $\Pcalbf^*,\Bcalbf^*$ and $\Hcalbf^*$ are involved, although they can be replaced by $\Pcalbf,\Bcalbf$ and $\Hcalbf$ due to Proposition~\ref{prop-2}.

The next proposition summarizes the two-level homogenization by the asymptotic analysis of the problem \eq{eq-mi5}; recall that therein all unknown functions and involved coefficients depend on both $\veps$ and $\dlt$ because of the domain partitioning \eq{eq-mi1}. Also note, that in \eq{eq-mi5}, $\pd_p\Om_f^\epsdlt$ vanishes, hence also  $\pd_p\Om = \emptyset$. By the consequence, in the boundary integral of the equilibrium equation \eq{eq-ma10-i}$_1$, the pressure $\bar p^0 = p^0$ is a part of the solution.



\begin{myproposition}{prop-3}
Solutions $(\ub^\dlt,\wb^\dlt,p^\dlt)$ of problem \eq{eq-me1} converge 
for $\dlt\rightarrow 0$ to the solutions of the macroscopic problem \eq{eq-ma10-i}.
\end{myproposition}

The proof relies on the convergence result \eq{eq-cnv2}. In Appendix~\ref{apx-4}, we concern its main part  which relates the expressions for the homogenized coefficients and the homogenized problem equations to the two-scale limit equation \eq{eq-ma1-i}.

\paragraph{Effective pressure in the mesoscopic porosity}
While $p^0$ is the macroscopic description of  the fluid pressure in the micropores, analogous quantity associated with mesoscopic porosity must be computed once the macroscopic problem has been solved. We define the average of $\hat p$,
\begin{equation}\label{eq-ma-P}
  \begin{split}
    P_c^0(x,t) & = \phi_c^{-1}\intY_{Z_c} \hat p(x,\cdot,t) = \Rcalbf: \eeb{\wb^0 + \dot\ub^0}  +  p^0\;,\\
    \mbox{ where }\quad \Rcal_{ij} & = \phi_c^{-1}\intY_{Z_c}\hat\vphi^{ij}\;, \quad \Rcal_{ij} = \Rcal_{ji}\;.
\end{split}
\end{equation}
This expression reveals that the mesoscopic pressure $P_c^0$ deviates from the micorpore pressure $p^0$ by the $\Rcalbf$-projection of the mesocopic fluid velocity deformation.

\paragraph{Differential equations --- strong formulation} Assuming enough regularity of all fields involved in \eq{eq-ma10-i}, obvious integration by parts  yields the following set of equations to be satisfied by the triple $(\ub^0,\wb^0,p^0)$,
\begin{equation}\label{eq-ma11-i}
\begin{split}
- \nabla\cdot\left(\Acalbf\eebx{\ub^0} - p^0\Bcalbf\right)
+ \hb^0 & = \fb^\blk\;,\\
 \Bcalbf:\eebx{\dot\ub^0} + \nabla\cdot\jb^0 +  \Mcal\dot p^0  & = 0\;, \\
 - \nabla\cdot\Scalbf\eebx{\wb^0 + \dot\ub^0}  - \hb^0 & = 0;,
\end{split}
\end{equation}
where the symmetry relations stated in Proposition~\ref{prop-2}, (i) have been applied , and
\begin{equation}\label{eq-ma12-i}
  \begin{split}
    \mbox{ rate of momentum  }\quad   \hb^0 & = -\Pcalbf^*(\nabla_x p^0 - \fb^f) + \Hcalbf^*\wb^0 \\
    & = -\Pcalbf^T(\nabla_x p^0 - \fb^f) - \Hcalbf\wb^0\;,\\
   \mbox{ seepage flow }\quad   \jb^0 & = -\Kcalbf(\nabla_x p^0 - \fb^f) - \Pcalbf\wb^0\;.
  \end{split}
\end{equation}
As the byproduct, the Neumann-type boundary conditions are obtained according to the expression derived when integrating by parts in \eq{eq-ma10-i} to get the strong formulation. These conditions can be specified through given $\gb,\bar p^0, \bar w_n$,
\begin{equation}\label{eq-ma13}
  \begin{split}
 \nb\cdot \sigmabf^\por = \nb\cdot\left(\Acalbf\eebx{\ub^0} - p^0\Bcalbf\right) & = -\bar\phi_c \bar p^0 \nb + \phi_m \bar\phi_s \gb\quad \mbox{ on } \pd_\sigma\Om\;,\\
 \nb\cdot\jb^0 & = - \bar\phi_m\bar\phi_f \bar w_n^\mic + \bar\phi_c \bar w_n^\meso\quad \mbox{ on } \pd_w\Om\;,\\
 \nb\cdot \sigmabf^\flow  = \nb\cdot\Scalbf\eebx{\wb^0 + \dot\ub^0}  & = \bar \phi_c \bar p^0 \nb \quad \mbox{ on } \pd_p\Om\;,
  \end{split}
\end{equation}
whereby the complementary Dirichlet conditions  on $\ub^0$, $\wb^0$ and $p^0$ are imposed,
\begin{equation}\label{eq-ma14}
  \begin{split}
    \ub^0 & = 0\quad \mbox{ on } \pd_u\Om\;,\\
    p^0 & = \bar p\quad \mbox{ on } \pd_p\Om\;,\\
    \wb^0 & = \bar\wb \quad \mbox{ on } \pd_u\Om\;.
  \end{split}
\end{equation}

\begin{myremark}{rem-BC}
\begin{enumerate}
\item The total (macroscopic) poroelastic stress $\sigmabf^\por = \Acalbf\eebx{\ub^0} - p^0\Bcalbf$ describes a static stress in the mixture of the fluid and solid associated with the microporosity represented by the matrix $\Om_m^\dlt$. The stress $\sigmabf^\flow$ represents the viscous stress associated with the flow in the primary porosity $\Om_c^\dlt$.
\item Vector $\hb$  associated with the fluid redistribution between the two porosities describes momentum interaction. The total equilibrium of the mixture is obtained upon summation \eq{eq-ma11-i}$_1$ and \eq{eq-ma11-i}$_3$, thus $\nabla\cdot( \sigmabf^\por + \sigmabf^\flow) + \fb^\blk = 0$.
\item If Proposition~\ref{prop-2}, (ii) applies (for an isotropic mesoscopic permeability in a periodic medium), $\hb = -\phi_c(\nabla_x p^0 - \fb^f) + \Hcalbf(\kappa(\nabla_x p^0 - \fb^f) -  \wb^0)$. It is worth noting, that  $\kappa(\nabla_x p^0 - \fb^f) -  \wb^0$ expresses a bulk velocity of the fluid in both the meso- and microporosities.
\item Instead of \eq{eq-ma14}$_3$, merely normal fluid velocity can be imposed, \ie $\wb\cdot\nb = \bar w_n$ on $\pd_w\Om$.
  Recall that $\pd_p\Om = \pd\Om \setminus \pd_w\Om$. 
\end{enumerate}
The boundary conditions for the model represented by equations \eq{eq-ma11-i} and \eq{eq-ma12-i} can be considered in a generalized setting. For instance, on normal components of the displacements can be prescribed, or mesoscopic pore pressure can be prescribed ``indirectly `` by virtue of \eq{eq-ma-P}.
\end{myremark}

\section{Numerical example}\label{sec-numex}

\begin{figure}[H]
	\centering
	\begin{subfigure}{0.45\linewidth}
		\includegraphics[width=\linewidth]{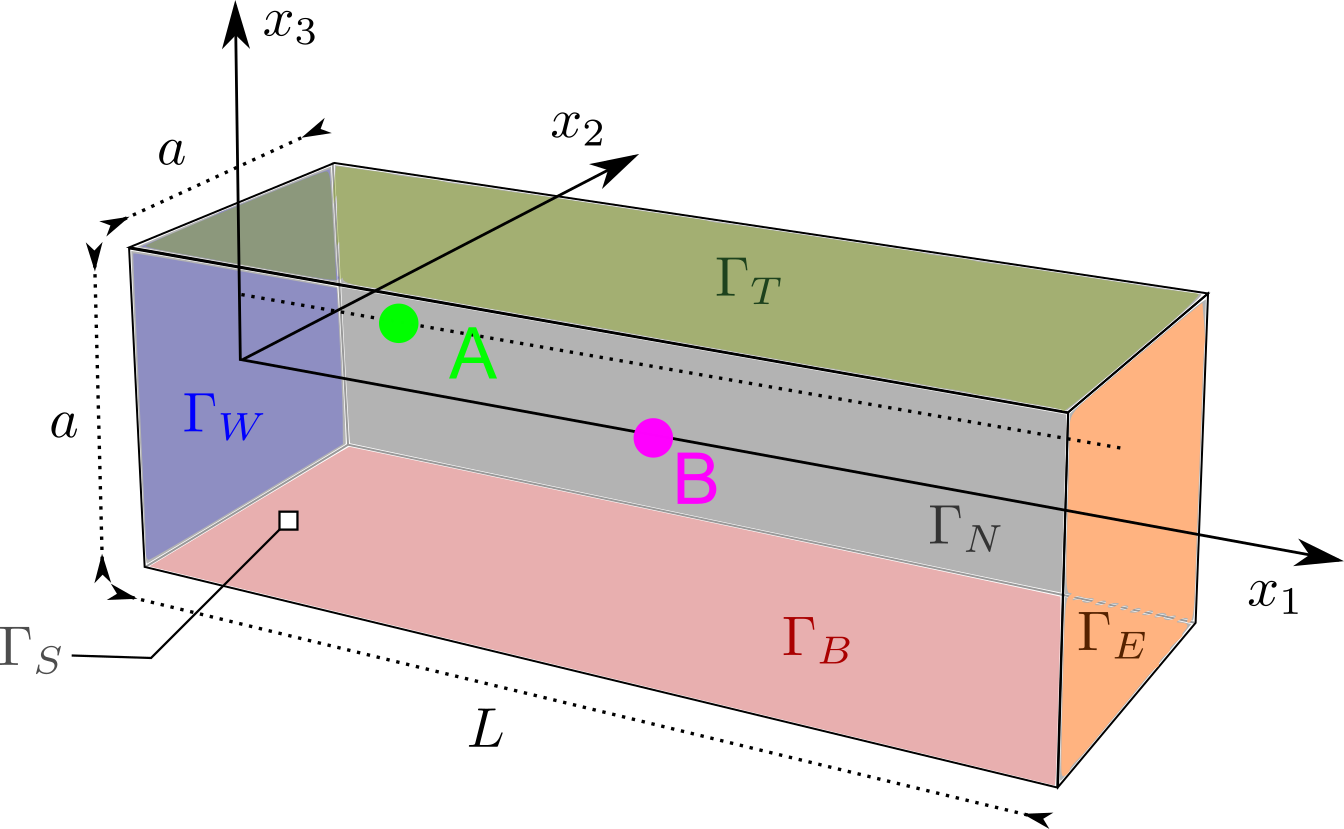}
		\caption{{Boundary decomposition and location of points $x^A$ and $x^B$ in the block.}}\label{fig-block}
	\end{subfigure}
	\begin{subfigure}{0.45\linewidth}
    \includegraphics[width=\linewidth]{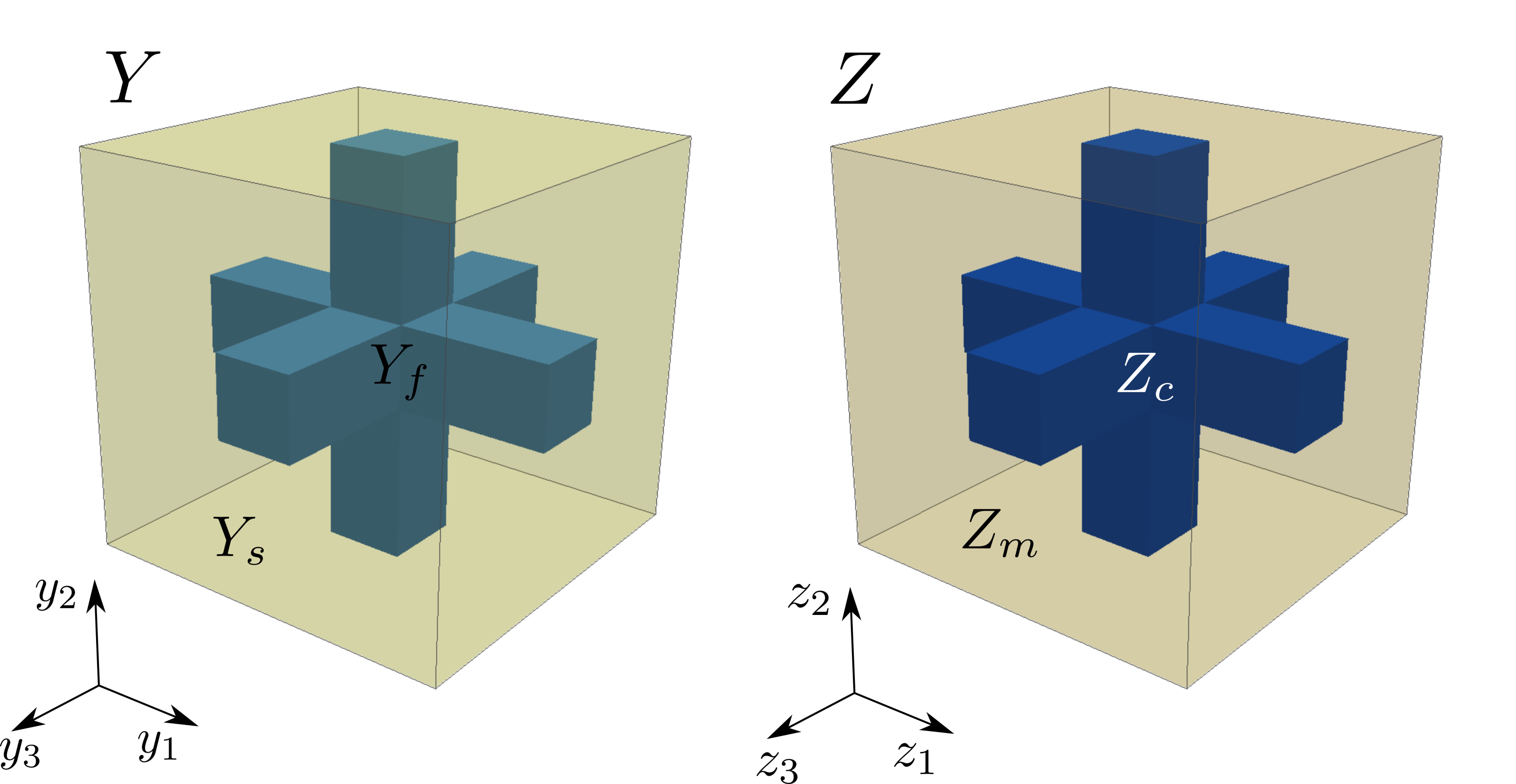}
    \caption{{Representative unit cells $Z$ and $Y$.}}\label{fig-cells-ZY}
	\end{subfigure}

\end{figure}

We present a numerical example of the hierarchical flows in a 3D deforming double-porosity structure described by the model proposed in this paper. The main purpose is to illustrate properties of this model in the context of the fluid redistribution between the two porosities induced by the deformation. To this aim we consider a simple 3D specimen shaped as a regular hexahedron, further called ``the block'' $\Om = ]0,L[\times]-a,a[\times]-a,a[$ with the dimensions $L = 10$~m and $a = 1.7$~m, see Fig.~\ref{fig-block}. The micro- and the mesoscopic porosity are generated by simple geometries defined by virtue of the decomposition of the micro- and mesoscopic periodic cells $Y$ and $Z$, respectively, see Fig.~\ref{fig-cells-ZY}.
    The material properties relevant to the microscopic scale are displayed in Table~\ref{tab-mat}. Since the 1st level upscaling is featured by the proportionality between the fluid viscosity and the pore size, \ie $\eta^\veps = \veps^2 \bar \eta_p$ and $\ell^\mic = \veps L$, the numerical model is established for a given size of the microstructure, ${\veps:=\veps_0 = 10^{-2}}$. 

\begin{table}[h]
    \centering
    \begin{tabular}{cccccc}
      Viscosity &Scale parameter&Young modulus &Poisson ratio & Pressure BC &Displacement BC \\
        \hline
        $\eta $& $\veps_0$&$E$& $\nu$& $\bar{p}^0$& $\bar{u}_n$\\
        $5\times 10^{-4}$&$10^{-2}$&$3\times 10^{9}$&$0.34$&$10^{4}$&$10^{-3}$\\
        Pa·s & - & Pa & - & Pa& m\\
        \hline
    \end{tabular}
    \caption{Parameters of the micromodel and of the prescribed boundary conditions (BC). Note that $\eta = \eta_c$ , while $\bar\eta_p = \eta/\veps_0^2$ is employed in the local problem \eq{eq-S3}.}\label{tab-mat}
\end{table}

\subsection{Boundary and initial conditions}
    
    To describe the boundary conditions, we refer to the faces of the block by the intuitive notation, such that $\Gamma_E$ stands for the ``east side'' with the normal aligned with $x_1$-axis, whereas the ``north side'' $\Gamma_N$ has its normal  aligned with $x_2$-axis. Then $\Gamma_T$ and  $\Gamma_B$ refer to the top and bottom sides, respectively, see Fig.~\ref{fig-block}.

    Though the example is arranged as a 3D test, our aim is to study the interactions in the double-porosity medium using a quasi one-dimensional test, such that the response can be assessed rather intuitively and gives a clear interpretation of the studied phenomenon. The fluid redistribution in time is driven by an unevenly applied deformation of the block whose mesoscopic pores on all faces are closed, \ie $w_n^\meso = 0$, which is respected by the surface porosity $\bar \phi_c = 0$. The micropores are open at the ``east'' and the ``west'' sides, where the pressure $p^0$ associated with micropores is prescribed. As the consequence this pressure also charges the solid surface through the traction forces $\gb$. Such a situation is specified by the following boundary conditions.

\paragraph{Boundary conditions}
\begin{list}{}{}
      
\item on $\Gamma_W$ (west side): $p^0=\bar p^0$, $\bar w_n^\meso = \wb\cdot\nb = 0$, $\gb =-\bar p^0\nb$,
\item on $\Gamma_E$ (east side): $p^0=\bar p^0$, $\bar w_n^\meso = \wb\cdot\nb = 0$, $\gb =-\bar p^0\nb$, 

\item on $\Gamma_N \cup \Gamma_S$ (north and south sides): $\bar w_n^\mic = 0$,  $\bar w_n^\meso = \wb\cdot\nb = 0$,  $\gb = \bmi{0}$,

\item on $\Gamma_T$ (top side): $\bar w_n^\mic = 0$,  $\bar w_n^\meso = \wb\cdot\nb = 0$,   $\ub\cdot\nb = \bar u_n(x-L/2)t$ and $\gb\cdot\tb = 0$,
          
\item on $\Gamma_B$ (bottom side): $\bar w_n^\mic = 0$,  $\bar w_n^\meso = \wb\cdot\nb = 0$,   $\ub = 0$.
\end{list}
Above $\tb$ is any tangent vector on the particular surface, such that on $\Gamma_T \cup \Gamma_S \cup \Gamma_N$ the medium can slide freely along the block faces.
Recall that $\bar w_n^\meso$ and $\bar w_n^\mic$ are the normal seepage velocities associated with the meso- and micro-pores, see \eq{eq-ma10-i}.

\paragraph{Initial conditions}
To initiate the simulation with a consistent initial conditions, we first consider the steady state of the problem \eq{eq-ma10-i}, where all time derivatives vanish. The boundary conditions are specified, as above with $t = 0$, such that $\ub\cdot\nb = 0$ on $\Gamma_T$. The displacement response $\ub^0(t=0)$, see Fig.~\ref{fig-e0}, is induced by the deformation caused by the pressure $\bar p^0$ acting on the ``east'' and ``west'' sides. The steady flow $\wb^0(t=0) = 0$ and pressure $p^0(t=0) = \bar p^0$. The fields $\ub^0(t=0)$ and $p^0(t=0)$ constitute the initial conditions.

\subsection{Remarks on the FE approximation}
The macroscopic model and all the local problems at the meso- and microscopic levels 
were implemented in our in-house developed software SfePy, \cite{cimrman_2014:sfepy} which allows for multiscale simulations using the FE method. The macroscopic fields $\ub^0$ and $\wb^0$ were approximated using the Lagrangian elements Q2, while the pressure field $p^0$ was approximated by the Q1 elements.

For the time discretization with the time step ${\Delta t = 0.025}$ the time derivatives $\dot \ub^0$ and $\dot p^0$ were approximated by the backward finite differences, such that the implicit time-integration scheme was employed.

\subsection{Results of the simulation}
The effective coefficients of the mesoscopic model are computed according to \eq{eq-h8}-\eq{eq-B1a}, so that the characteristic responses \eq{eq-me13a} and \eq{eq-me15b-i} of the periodic mesoscopic structure can be resolved. Consequently, the effective coefficients of the macroscopic model are evaluated using \eq{eq-ma2-i}-\eq{eq-ma4-i}. The initial state at $t = 0$ is computed first, then the evolutionary problem  \eq{eq-ma10-i} is solved within a give time interval.  

\begin{figure}[H]
	\centering
	\begin{subfigure}{0.45\linewidth}
	\includegraphics[width=\linewidth]{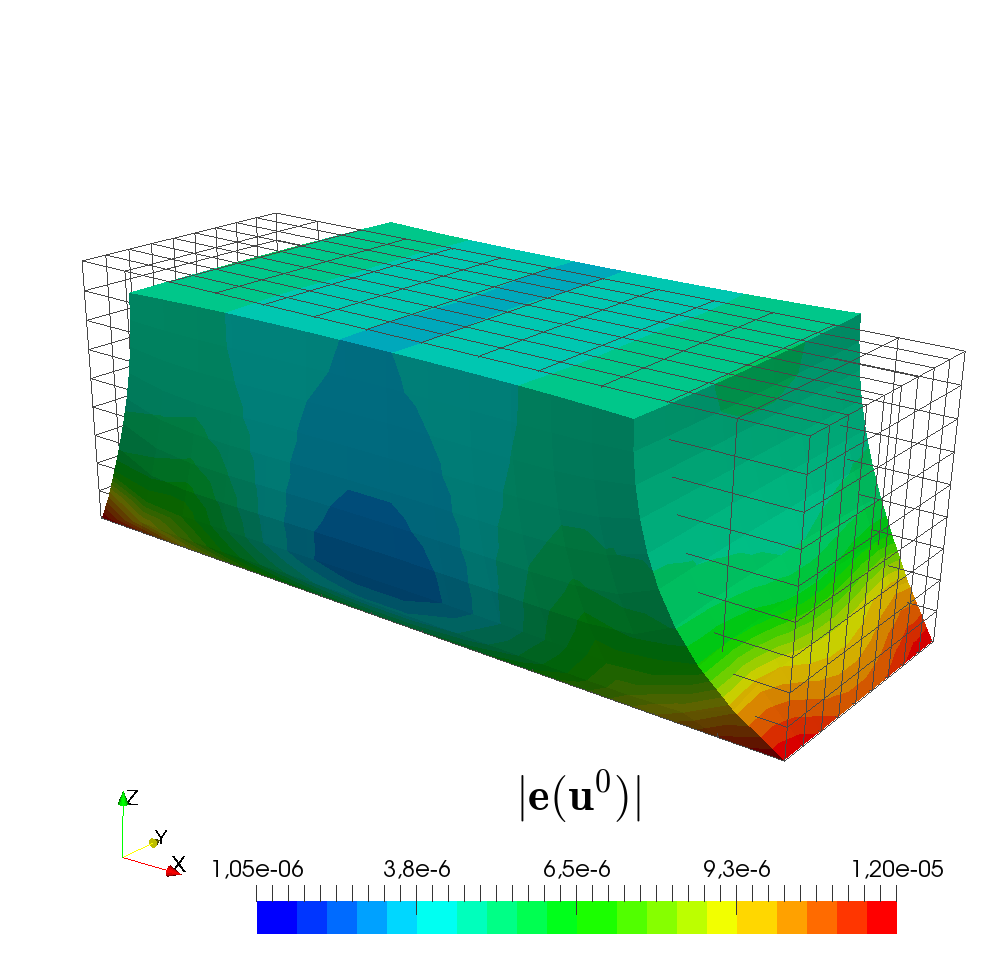}
	\caption{Strain at $t=0$s.}\label{fig-e0}
	\end{subfigure}\hfill
	\begin{subfigure}{0.45\linewidth}
	\includegraphics[width=\linewidth]{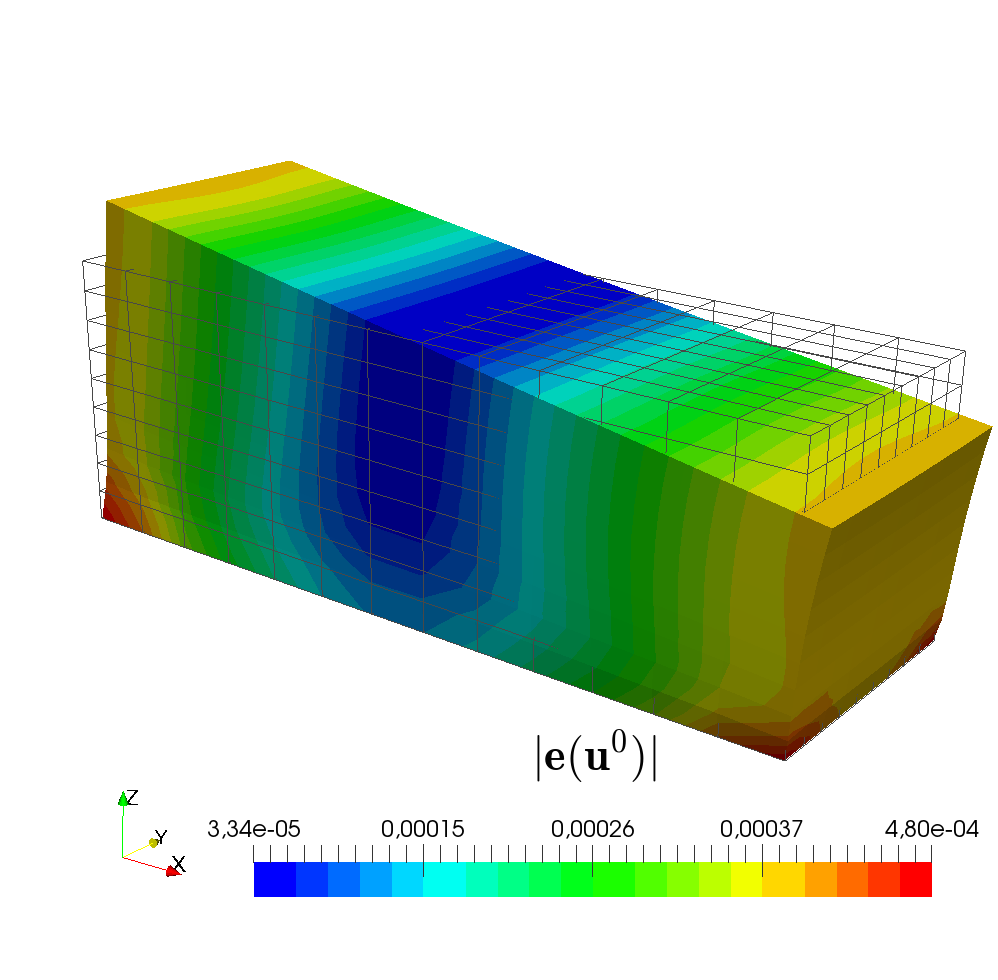}
	\caption{Strain at $t=1$s.}\label{fig-e}
	\end{subfigure}
        \caption{Macroscopic strains; norm $|\eeb{\ub^0}| = (e_{ij}(\ub^0)e_{ij}(\ub^0))^{1/2}$ displayed in the deformed configuration with scaled displacements: (a) 5000$\times$, (b) 3000 $\times$.}\label{fig-strain}
\end{figure}

To illustrate distributions of the deformation, pressure and flow fields, we depict the model response at time $t = 1$~s, however, as discussed below, the spatial distributions of the fluid flow and pressures attain the same patterns for $t > 2$ ms.  In Figs.~\ref{fig-e0} and \ref{fig-e}, 
strains $|\eeb{\ub^0}|(x)$  are depicted in the deformed configuration; Fig.~\ref{fig-e0} shows the initial strain at $t=0$ which is induced by the steady pressures $P_c^0=p^0$.

\begin{figure}[H]        
\begin{subfigure}{0.45\linewidth}
	\includegraphics[width=\linewidth]{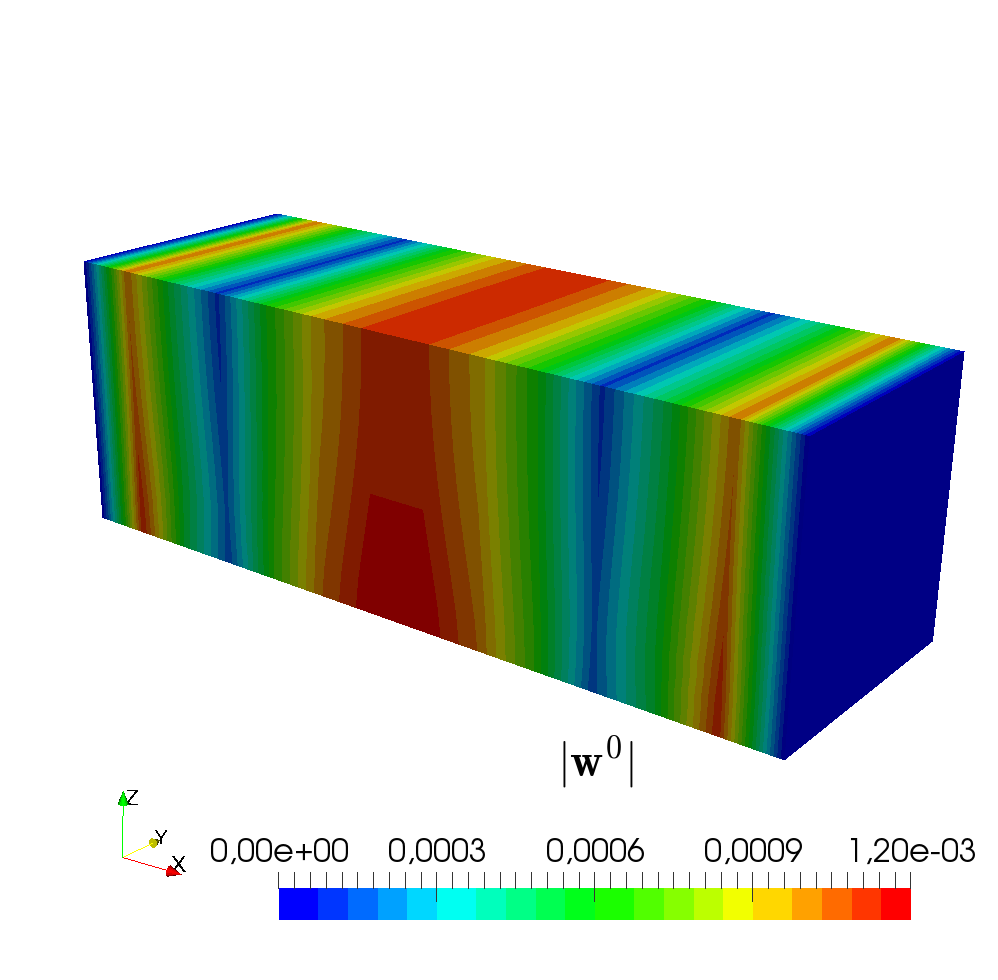}
	\caption{Velocity field magnitudes $|\wb^0|$.}\label{fig-wm}
 \end{subfigure} \hfill
\begin{subfigure}{0.45\linewidth}
	\includegraphics[width=\linewidth]{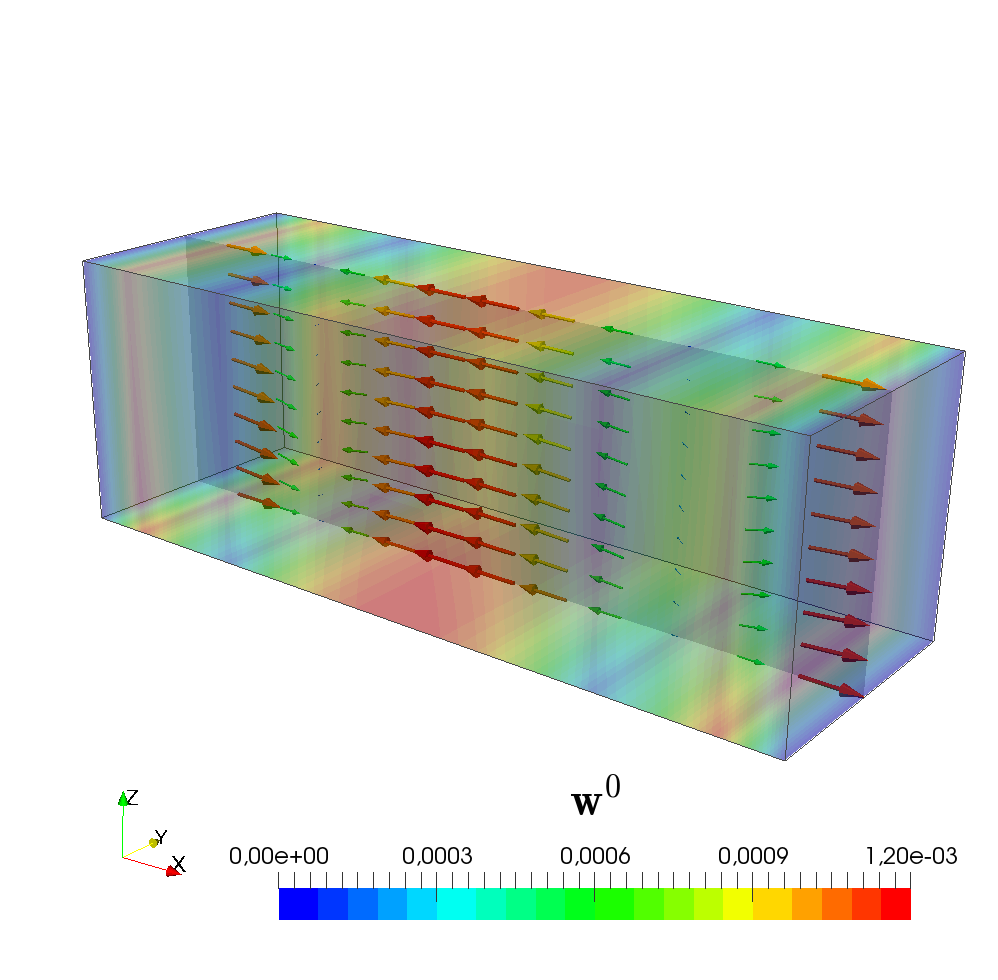}
	\caption{Velocity field $\wb^0$. (Magnitudes correspond to the arrow lengths).}\label{fig-w} 
 
\end{subfigure}
	\begin{subfigure}{0.45\linewidth}
	\includegraphics[width=\linewidth]{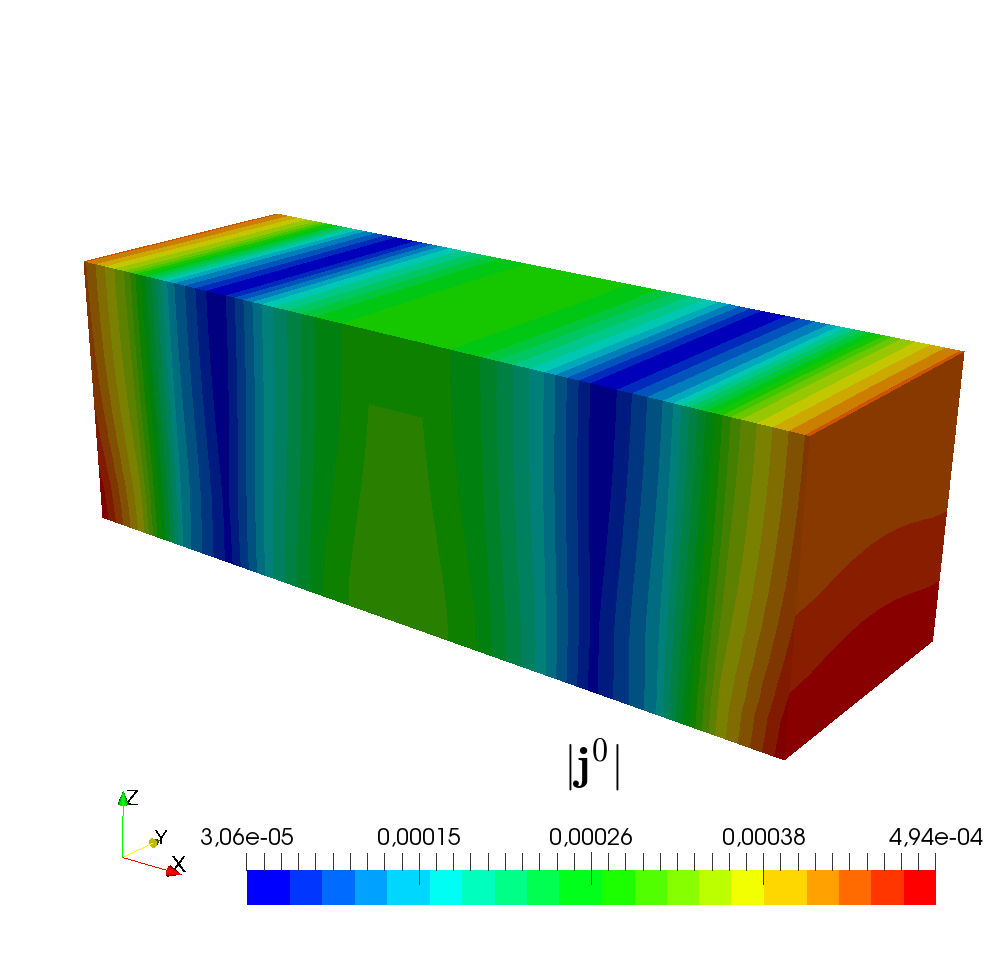}
	\caption{Seepage flow field magnitudes $|\jb^0|$.} \label{fig-jm}
 	\end{subfigure} \hfill
	\begin{subfigure}{0.45\linewidth}
	\includegraphics[width=\linewidth]{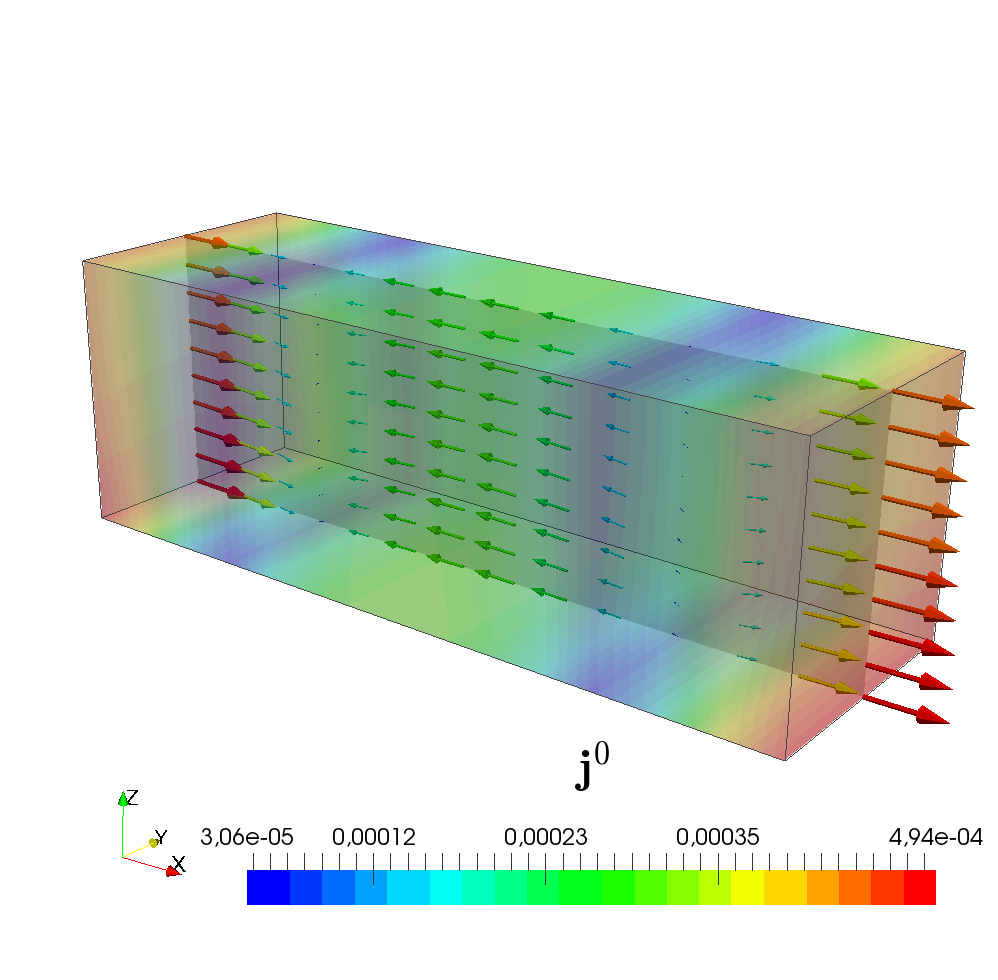}
	\caption{Seepage flow field $\jb^0$. (Magnitudes correspond to the arrow lengths).}
 \label{fig-j}
 	\end{subfigure}

        \caption{Fluid flow in the double-porosity structure at the macroscopic level.} 
\end{figure}

\begin{figure}[H]        
\begin{subfigure}{0.45\linewidth}
	\includegraphics[width=\linewidth]{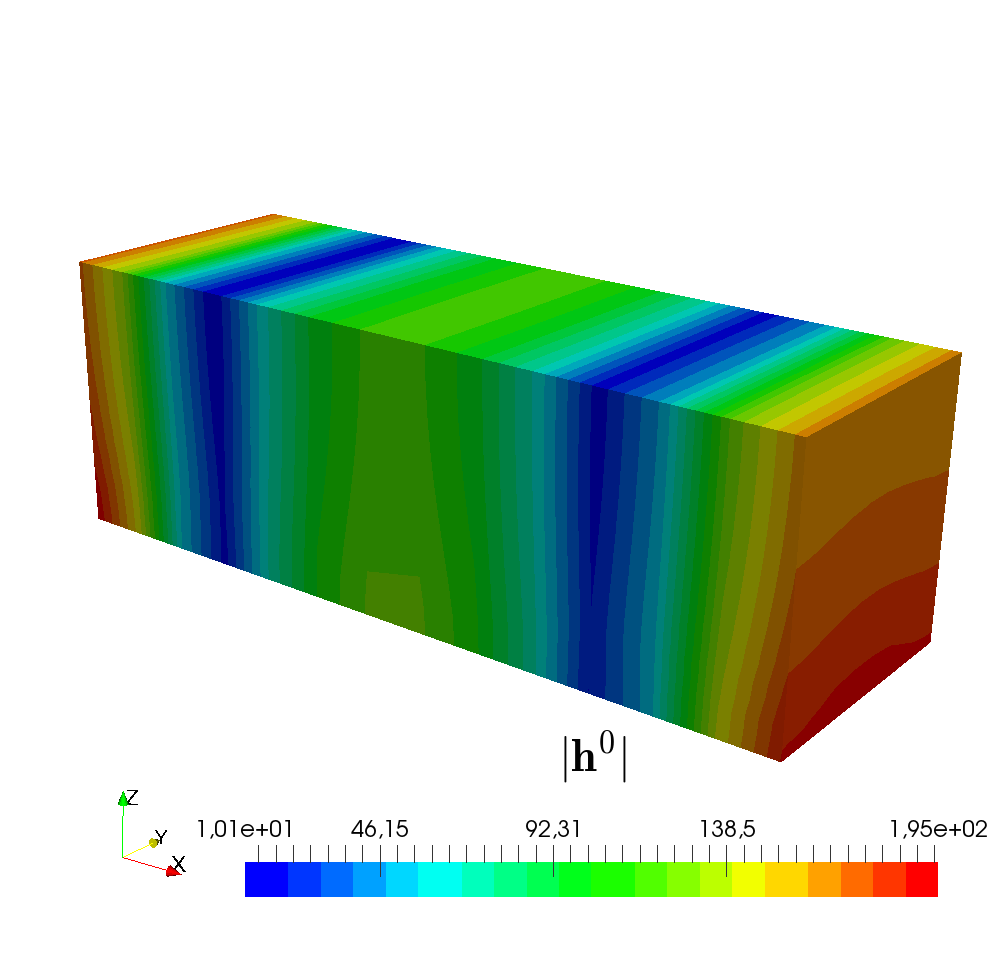}
	\caption{Magnitudes of momentum exchange $|\hb^0|$.} \label{fig-hm}
 	\end{subfigure} \hfill
	\begin{subfigure}{0.45\linewidth}
	\includegraphics[width=\linewidth]{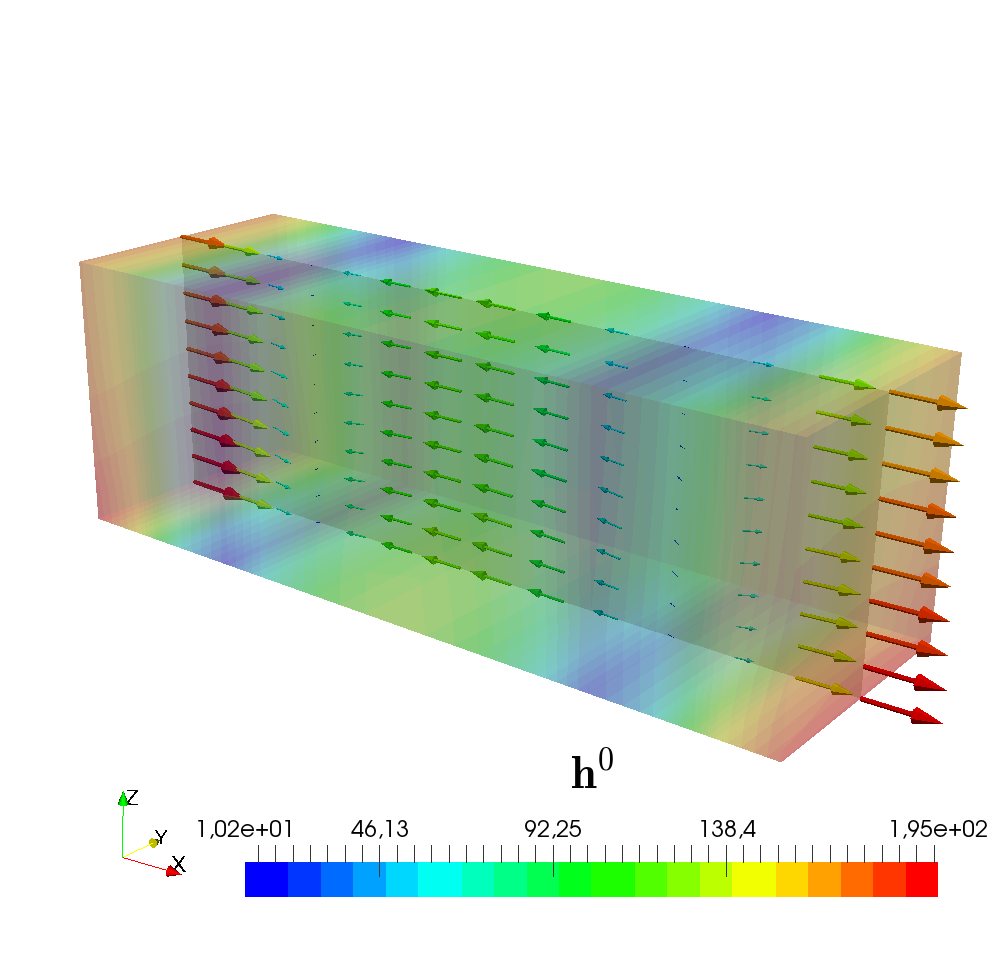} 
	\caption{Momentum exchange $\hb^0$. (Magnitudes correspond to the arrow lengths).}
 \label{fig-h}
 	\end{subfigure}
        
        \caption{Momentum exchange $\hb^0$ between the two porosities.} \label{fig-hh}
\end{figure}

The flows in the hierarchical porosity are illustrated in Figs.~\ref{fig-wm},\ref{fig-w} and \ref{fig-jm},\ref{fig-j}, in the latter the total macroscopic seepage flow $\jb^0$ is computed by \eq{eq-ma12-i}. It is worth noting that the macroscopic flow $\wb^0$ in the channels (the primary porosity) fluctuates with ``axial'' position according to $x_1$, see  Fig.~\ref{fig-wm}, whereas magnitudes of the overall macroscopic seepage are more evenly distributed, see Fig.~\ref{fig-jm}. The directions of the two flow fields are displayed in Figs.~\ref{fig-w} and \ref{fig-j}. Due to the ``nonsymmetric'' deformation (assumed symmetry plane at $x_1 = L/2$ with the normal aligned with $x_1$-axis), the flow fields are nonsymmetric as well. There are two flow-stagnation planes where the divergences $\nabla\cdot\jb^0$ and $\nabla\cdot\wb^0$ have opposite signs. Similar distribution of flux $\hb^0$ is shown in Figs.~\ref{fig-hm} and \ref{fig-h}.

\begin{figure}[H]        
\begin{subfigure}{0.45\linewidth}
	\includegraphics[width=\linewidth]{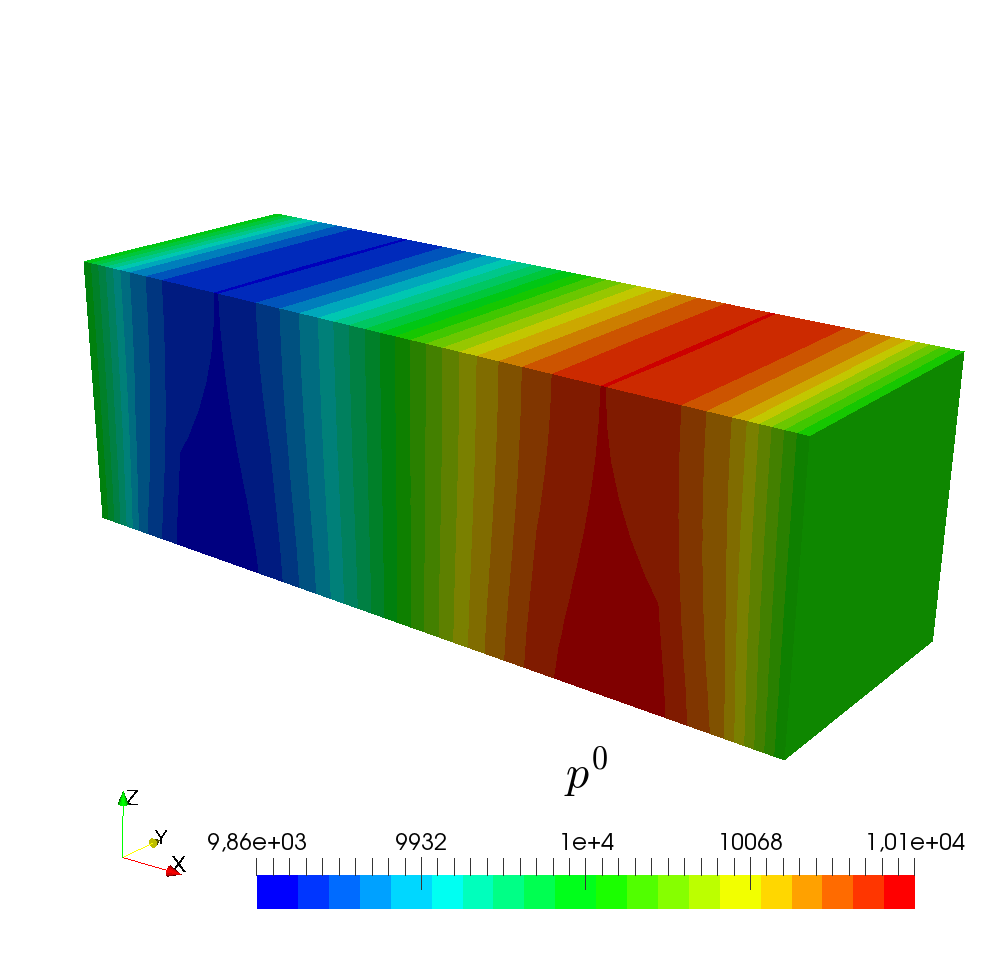}
	\caption{Pressure $p^0$ (fluid in microporosity).}\label{fig-p0}
 \end{subfigure} \hfill
\begin{subfigure}{0.45\linewidth}
	\includegraphics[width=\linewidth]{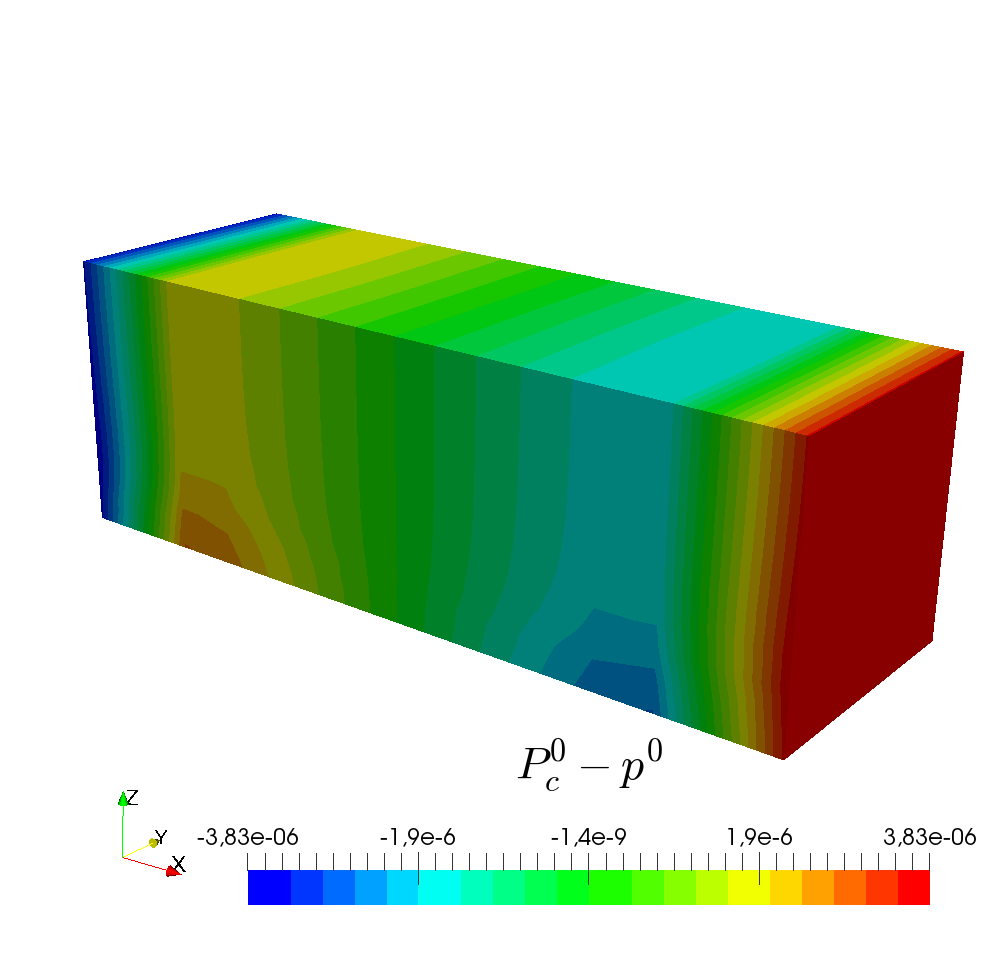} 
	\caption{Difference $P_c^0-p^0$ (fluid in mesoscopic channels).}\label{fig-Pc-p0} 
 
\end{subfigure}

  \caption{Fluid pressure fields associated with the micro- and meso-pores. Note that $P_c^0-p^0$ captures effects of the mesoscopic flow divergence.}\label{fig-Pp}
           
\end{figure}

The pressure distribution $p^0$ and the difference $P_c^0-p^0$ are displayed  in Figs.~\ref{fig-p0} and \ref{fig-Pc-p0}. 
According to \eq{eq-ma-P}, pressure $P_c^0$ deviates form $p^0$ only due to the flow deformation $\eeb{\wb^0 + \dot\ub^0}$; due to the micro- and mesoscopic structure, both symmetric, tensor $\Rcalbf$ is isotropic, so that  this test captures the influence of $\nabla\cdot(\wb^0 + \dot\ub^0)$.

\begin{figure}[H]        
\begin{subfigure}{0.45\linewidth}
	\includegraphics[width=\linewidth]{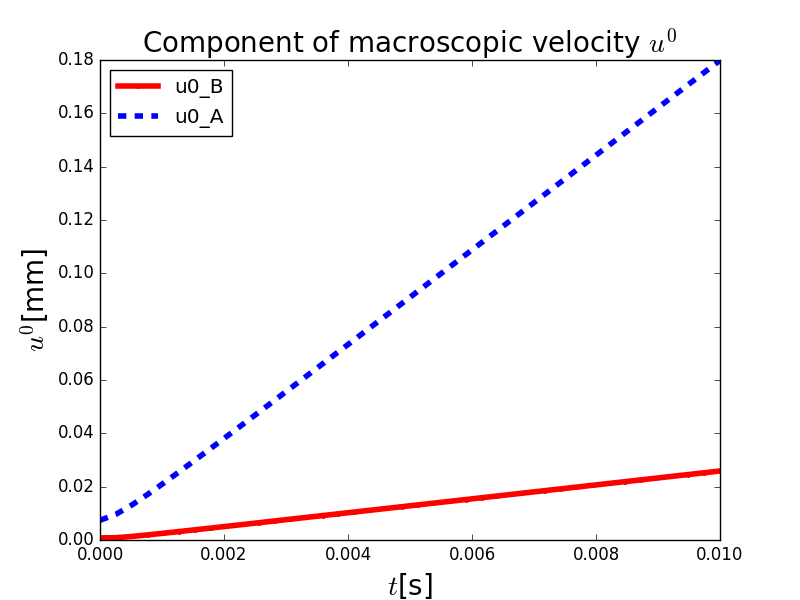}
	\caption{Displacement magnitude $u^0$.}\label{fig-ut}
 \end{subfigure} \hfill
\begin{subfigure}{0.45\linewidth}
	\includegraphics[width=\linewidth]{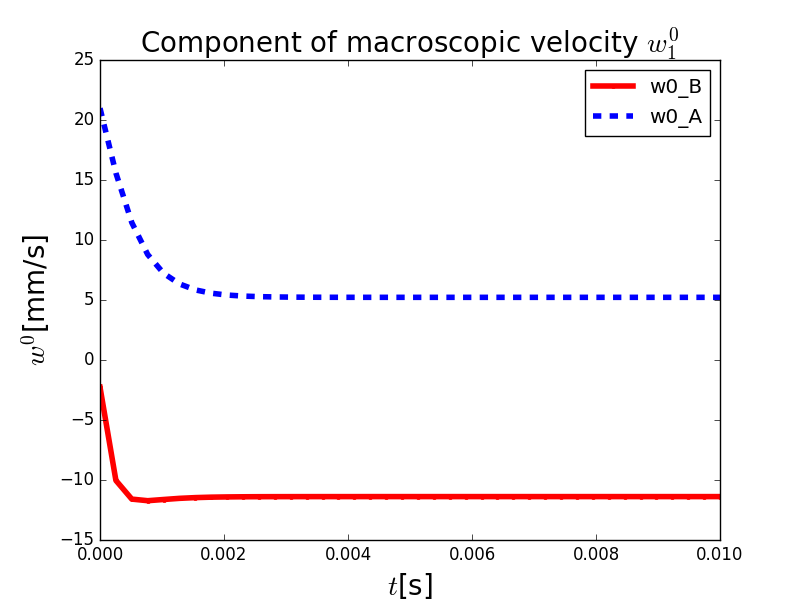}
	\caption{Velocity $w_1^0$ (flow in mesoscopic channels).}\label{fig-wt} 
\end{subfigure}

\begin{subfigure}{0.45\linewidth}
  \includegraphics[width=\linewidth]{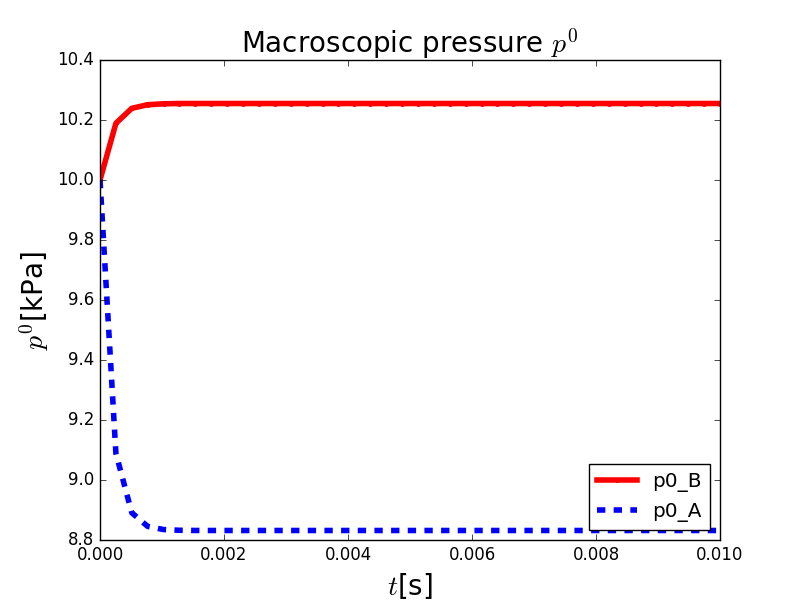}
	\caption{Pressure $p^0$ (microporosity)}\label{fig-p0t}
 \end{subfigure} \hfill
\begin{subfigure}{0.45\linewidth}
	\includegraphics[width=\linewidth]{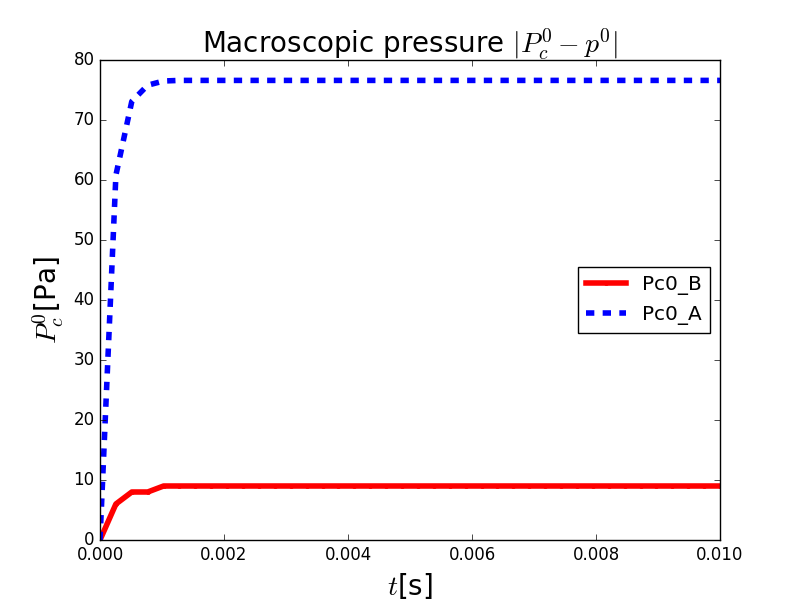}
	\caption{Difference $P_c^0-p^0$.}\label{fig-Pc0-p0-t} 
\end{subfigure}

        \caption{Evolution of state quantities at points  $x^A$ and $x^B$.}\label{fig-response-t}
         
\end{figure}

Finally, Fig.~\ref{fig-response-t} shows the macroscopic response at two points, $x^A$ and $x^B$ whose locations are marked in Fig.~\ref{fig-block}. While the displacement increases as a  linear function of $t$, the pressures and seepage velocities tend to constant values, so that a new steady flow in the deforming specimen is established. This effect is certainly due to the model linearity, since the deformation and other state quantities do not influence the effective medium coefficients. 
The fluid redistribution between the two porosities influenced by the deformation-dependent permeability, see \eg \cite{Rohan-AMC,Sandstrom-Larssen-CMAME2016}, and other coefficients would produce time-dependent flow patterns in the block, such that the above mentioned steady state would not be achieved.

\section{Conclusion and discussion}\label{sec-concl}
In this paper we proposed a new model of viscous fluid flow in the double porosity structure with elastic deformable skeleton. The model was obtained by two-level homogenization with two independent parameters, $\veps$ and $\delta$, which are associated with the microporosity and the mesoporosity characteristic lengths, respectively. Due to the scaling ansatz for the viscosity in the microchannels, the first level asymptotic analysis \wrt $\veps\rightarrow 0$ leads to the Biot-Darcy model describing the mesoscopic porous matrix. The fluid is redistributed between the matrix and the channels, where the flow is governed by the Stokes flow. The second level upscaling with $\delta\rightarrow 0$ results in the macroscopic model which involves the three macroscopic fields: the displacements of the solid phase, the relative velocity of the fluid \wrt solid (the seepage velocity) representing the flow in the mesoscopic channels, and the pressure associated with
the micropores. The mesoscopic flow is governed by the Brinkman model which is coupled with the Biot-Darcy model arising from the same kind of the model relevant to the mesoscopic level, although with modified poroelasticity and permeability coefficients. Due to the interaction between the two media at the mesoscopic level, \ie the matrix and the channels, the macroscopic model involves coupling tensors which reveals the Onsager reciprocity principles.
Interestingly, these coefficients appear already in the rigid skeleton model, but now enter also the equilibrium equation relevant  to the model with deformable skeleton only. 

The present work is a natural continuation of the previous research reported in \cite{Rohan-etal-CaS2017}, where the same kind of the double porosity media was treated, however, with rigid skeleton and assuming steady state incompressible flows only, leading to the Darcy-Brinkman model of steady flows. In the present paper we consider quasistatic events, such that the inertia effects are neglected. The fluid compressibility is considered to obtain the two scale equations of the macroscopic model, however, the scale decoupling procedure in the second level homogenization is presented for an incompressible fluid only. The derived macroscopic model contains the above mentioned Darcy-Brinkman model as its subpart.

There are several issues to be handled in a future research. The most striking question concerns with the mesoscopic interface conditions coupling the flow in the porous matrix and in the channels. On one hand, it appears that without any special treatment, pursuing the approach proposed in  \cite{CDG-Stokes-2005}, the free slip flow condition is obtained. On the other hand, the well-known Beavers-Joseph condition which links the tangential velocity jump with the viscous stress in the bulk fluid of the channels, can be obtained by a special treatment of the boundary layer of the interface, \cf \cite{Jamet-TiPM2009,Chandesris-TiPM2009,Jaeger-Mikelic-2009,Czochra-Mikelic-MMS-2012}. In any case, the interface conditions for flow pertain an open question namely when curved surfaces are considered.

To study the fully dynamic behaviour of this kind of the double porosity structure, it is important to account for inertia effects related to both solid and fluid phases. It will lead to a similar difficulty as the one arising from the fluid compressibility. To decouple the scales, the Laplace transformation in time is required, which introduces time convolution integrals in the upscaled models \cite{rohan-etal-jmps2012-bone}. In the context of the two-level upscaling procedure, this will complicate the homogenization procedure significantly.

As the further issue, the nonlinearity arising with large deformation, or material nonlinearity should be studied. For the one-level homogenization, see \eg \cite{Sandstrom-Larssen-CMAME2016}, the computational procedure leads typically to the ``FE-square'' complexity, which seems to be intractable in the context of the reiterated, or multi-step homogenization. For situations with moderate deformations, the phenomenon of deformation-dependent homogenized coefficients can be handled efficiently using the sensitivity approach, as proposed in \cite{Rohan-AMC}.

\paragraph{Acknowledgment}
This research is supported by project GACR 16-03823S of the Scientific Foundation of the Czech Republic and
 from European Regional Development Fund-Project ``Application of Modern Technologies in Medicine and Industry'' (No.~CZ.02.1.01/0.0/0.0/17~048/0007280).
Jana Turjanicov\'a is grateful to the project
 LO 1506 of the Czech Ministry of Education, Youth and Sports.

\begin{appendices}

\section{Notation and functional spaces}\label{apx-1}
\label{sec-Appendix-A} In the paper, the following notations are used.
\begin{list}{$\bullet$}{}




\item By $\pd_i = \pd_i^x$ we abbreviate the partial derivative $\pd/\pd x_i$.
    We use $\nabla_x = (\pd_i^x)$ and $\nabla_y = (\pd_i^y)$ when differentiation \wrt coordinate $x$ and $y$ is used, respectively.
    The symmetric gradient of a vectorial function
    $\ub$, $\eeb{\ub} = 1/2[(\nabla\ub)^T + \nabla\ub]$ where the transpose
    operator is indicated by the superscript ${}^T$.

\item The space $C^r(D)$ of $r$-times continuously differentiable functions on $D$.
The space $C_0^\infty(D)$ of $C^\infty(D)$ functions with the compact support in $D$, \ie vanishing on $\pd D$ including all their derivatives.

\item The Lebesgue spaces $L^2(D)$ of square-integrable functions on $D$. The
    Sobolev space $H^1(D)$ of the square-integrable functions up to the 1st order generalized derivative. The notation with bold and
    non-bold letters, \ie like $H^1(D)$ and $\Hdb(D)$ is used to distinguish
    between spaces of scalar, or vector-valued functions. 

\item The space $\Hpdb(D)$ is the Sobolev space of functions defined in $D$,
    integrable up to the 1st order of the generalized derivative, and which are
    Y-periodic in $\Om$. Moreover, we need the space $\Hpdbav(D) = \{\vb \in
    \Hpdb(D)\;|\;\int_{D} \vb = 0\}$.


\end{list}{}{}

\subsection{Highlights on the unfolding operator method}\label{apx-uf}
For the reader's convenience we recall the notion of the \emph{periodic
  unfolding method} and of the {\it periodic unfolding operator}, in
particular. We shall use the convergence results in the unfolded
domains $\Om \times Y$, or  which can be found in  \cite{Cioranescu-etal-2008}.

In order to define an extension  operator (from the channels to the matrix, or briefly  an ``off-channels'' extension), we
introduce the domain containing    the ``entire'' periods $\veps Y$:
\begin{equation*}\label{eq:3}
\begin{split}
\hat \Om^\veps & = \mbox{interior} \bigcup_{\zeta \in \Xi^\veps} Y_\zeta^\veps\;, \quad Y_\zeta^\veps= \veps (\ol{Y} + \zeta )\\
\mbox{ where } \Xi^\veps & = \{\zeta \in \ZZ^3\,|\; \veps (\ol{Y} + \zeta) \subset \Om\}\;.
\end{split}
\end{equation*}

For all $z \in \RR^3$,  let $[z]$ be the unique integer such that $z- [z] \in Y$.
We may write $z = [z]+\{z\}$ for all $z\in \RR^3$, so that
for all $\veps >0$, we get the unique decomposition
\begin{equation}\label{eq:3a}
x = \veps\left ( \left [\frac{x}{\veps}\right ] + \left \{\frac{x}{\veps}\right \}\right) =  \xi + \veps y
\quad \forall x \in \RR^3\;,\quad \xi = \veps\left [\frac{x}{\veps}\right ]\;.
\end{equation}
Based on this decomposition, the periodic unfolding operator
$\Tuftxt{}:  L^2(\Om;\RR) \rightarrow L^2(\Om \times Y;\RR)$ is defined as follows: for
 any function $v \in L^1(\Om;\RR)$, extended to $L^1(\RR^3;\RR)$ by zero outside $\Om$,
i.e. $v=0$ in $\RR^3 \setminus \Om$,
\begin{equation*}
\Tuf{v}(x,y) =
\left \{
\begin{array}{ll}
v\left( \veps \displaystyle  \left [\frac{x}{\veps}\right ] + \veps y \right)\;,
\quad &  x \in \hat\Om^\veps, y \in Y\;, \\
0 & \mbox{ otherwise }. \\
\end{array}
\right .
\end{equation*}
For product of any $u$ and $v$ the unfolding yields $\Tuf{uv}=\Tuf{u}\Tuf{v}$.

The following integration formula holds:
\begin{equation*}
\int_{\hat\Om^\veps} v\,dx = \frac{1}{|Y|}\int_{\Om \times Y} \Tuf{v}\,dy\,dx\quad
\forall v \in L^1(\Om)\;.
\end{equation*}
By $\Mcal_Y(\cdot)$ we denote the average operator over $Y$, if $\Tuf{w^\veps}\rightharpoonup \hat w$ weakly in $L^p(\Om\times Y)$, then ${w^\veps}\rightharpoonup \Mcal_Y(\hat w)$
weakly in $L^p(\Om)$.
For any $D\subset Y$,
$\intYsmall_{D} = \frac{1}{|Y|}\int_{D}$; the analogical notation is employed for any $A\subset Z$, thus $\intYsmall_{A} = \frac{1}{|Z|}\int_{A}$. Further, for any $D\subset Y$, $\Hpdb(D)$ is the Sobolev space $\Wb^{1,2}(Y) = \Hdb(Y)$ of vector-valued Y-periodic functions (indicated by the subscript $\#$).

\begin{myremark}{rem-Tuf}
In this paper we use two unfolding operators: operator $\Tuf{~}$ is employed in the first level homogenization reported in Section~\ref{sec-1-homog}, to upscale the microporosity in $\Om_1^\epsdlt$ when
$\veps\rightarrow 0$. Then in Section~\ref{sec-2-homog} we introduce
operator $\Tufd{~}$ which is employed in the second level homogenization when upscaling the mesoscopic structure; in the context of decomposition \eq{eq:3a}, we shall consider $\{x/\dlt\}_Z = z \in Z$. 
\end{myremark}


\section{1st level upscaling}\label{apx-2}
Consider the interfacial stress working,
\begin{equation}\label{eq-auxmic1}
  \begin{split}
    \int_{\Gamma_{sf}^\veps} \sigmabf^{f,\veps}:\nb^\sx\otimes\vb^\veps =
    \int_{\Gamma_{sf}^\veps} (\veps^2\bar\eta_p (\eeb{\wb^{\veps} + \dot{\tilde\ub}^{\veps}}) - p^\veps \Ib):\nb^\sx\otimes\vb^\veps \\
    =\int_{\Om_m} \frac{1}{\veps}\intY_{\Gamma}\left((\veps^2\bar\eta_p\Tuf{\eeb{\wb^{\veps} + \dot{\tilde\ub}^{\veps}}} -\Tuf{p^\veps}\Ib \right):\nb^\sx\otimes(\vb^0 +\veps \vb^1)\\
    =\int_{\Om_m} \frac{1}{\veps}\intY_{\Gamma}\left((\veps^2\bar\eta_p
    \eeb{\wb^{\veps,0} + \ub^{\veps,0} + \veps\ub^{\veps,1}} - p^{\veps,0} - \veps p^{\veps,1}\right):\nb^\sx\otimes(\vb^0 +\veps \vb^1)\;.
\end{split}
\end{equation}
The following convergences hold:
\begin{equation}\label{eq-auxmic2}
  \begin{split}
    \int_{\Om_m} \frac{1}{\veps}\intY_{\Gamma}\left(p^{\veps,0} + \veps p^{\veps,1}
    \right):\nb^\sx\otimes(\vb^0 +\veps \vb^1)\rightarrow
    \int_{\Om_m}\intY_{\Gamma} (p_m^0 \vb^1 + p_m^1 \vb^0)\cdot\nb^\sx\;,\\
    \int_{\Om_m} \frac{1}{\veps}\intY_{\Gamma}\veps^2\bar\eta_p\eeb{\Tuf{\ub^{\veps,0} + \veps\ub^{\veps,1}}}:\nb^\sx\otimes(\vb^0 +\veps \vb^1)\rightarrow 0\;,\\
    \int_{\Om_m} \frac{1}{\veps}\intY_{\Gamma}\veps^2\bar\eta_p(\frac{1}{\veps}\eeby{\wb^{\veps,0}}+\eebx{\wb^{\veps,0}}):\nb^\sx\otimes(\vb^0 +\veps \vb^1)\rightarrow
    \int_{\Om_m} \intY_{\Gamma}\bar\eta_p\eeby{\wb^0}:\nb^\sx\otimes\vb^0\;,
\end{split}
\end{equation}
so that
\begin{equation}\label{eq-auxmic3}
  \begin{split}
    \int_{\Gamma_{sf}^\veps} \sigmabf^{f,\veps}:\nb^\sx\otimes\vb^\veps \rightarrow
    \int_{\Om_m} \intY_{\Gamma}\left( \bar\eta_p\eeby{\wb^0}:\nb^\sx\otimes\vb^0
    - (p_m^0 \vb^1 + p_m^1 \vb^0)\cdot\nb^\sx\right)\;,
\end{split}
\end{equation}

For the mesoscopic problem we take $\vb^1 \equiv 0$ and the remaining interface integral \eq{eq-auxmic3} involving $\vb^0$ can be rewritten using the split \eq{eq-mi9},
\begin{equation}\label{eq-auxmic4}
  \begin{split}
 \intY_{\Gamma}\left( \bar\eta_p\eeby{\wb^0}:\nb^\sx\otimes\vb^0
    -  p_m^1 \vb^0)\cdot\nb^\sx\right) = \vb^0\phi_f \cdot (\nabla p_m^0 - \fb^f)\;.
\end{split}
\end{equation}
This follows from the strong form of the micro-problem \eq{eq-S3} which states the identity
$-\bar\eta_p \nabla_y^2\psibf^k + \nabla_y \pi^k = - \oneb^k$.

\section{2nd level upscaling}\label{apx-3}

\subsection{Limit expressions of the interface integrals}
Using the convergence result \eq{eq-cnv2} we inpect convergence of the interface integrals involved in \eq{eq-me1}, upon substituting the expansions \eq{eq-me2}.
\begin{equation}\label{eq-me3a}
\begin{split}
-\int_{\Gamma_{cm}^\dlt} p_m^\dlt \nb^\cx \cdot  \vb^\dlt = \int_{\Gamma_{cm}^\dlt} p_m^\dlt \nb^\mx \cdot  \vb^\dlt = 
\int_{\Om_m} \nabla\cdot(\tilde p_m^\dlt \vb^\dlt) - \int_{\pd_\ext\Om_m} p_m^\dlt \nb^\mx \cdot  \vb^\dlt\\
\rightarrow \int_\Om \intY_{Z_m}(\nabla_x p^0 + \nabla_z p^1)\cdot\vb^0 + \int_\Om\intY_{Z_m} p^0 \left( \nabla_x\cdot \vb^0 +  \nabla_z\cdot\vb^1\right)
- \int_{\pd\Om} \bar \phi_m p^0 \nb\cdot  \vb^0\\
= \int_\Om \intY_{Z_m} \nabla_z{p}^1\cdot\vb^0 + 
 \int_\Om \intY_{Z_m}{p}^0 \nabla_z\cdot\vb^1 
 + \int_{\pd\Om} (\phi_m - \bar\phi_m) {p}^0 \nb\cdot\vb^0 -\int_\Om {p}^0 \vb^0 \nabla_x \cdot\phi_m\\
 = \int_\Om \vb^0\cdot \intY_{Z_m} \nabla_z{p}^1 + \int_\Om p^0 \int_{\Gamma_Z} \nb^\mx \cdot\vb^1 -\int_\Om \phi_c\nabla_x \cdot(p^0 \vb^0) + \int_{\pd\Om} \bar\phi_c {p}^0 \nb\cdot\vb^0
 \;.
\end{split}
\end{equation}
Above the following identity has been applied:
\begin{equation}\label{eq-me3a-i}
\begin{split}
  \int_{\pd\Om} (\phi_m - \bar\phi_m) {p}^0 \nb\cdot\vb^0 -\int_\Om {p}^0 \vb^0 \nabla_x \cdot\phi_m
 & = \int_{\pd\Om} (\bar\phi_c - \phi_c){p}^0 \nb\cdot\vb^0 + \int_\Om {p}^0 \vb^0 \nabla_x \cdot\phi_c\\
 & = -\int_\Om \phi_c\nabla_x \cdot(p^0 \vb^0) + \int_{\pd\Om} \bar\phi_c {p}^0 \nb\cdot\vb^0
  \;.
\end{split}
\end{equation}

To pass to the limit in \eq{eq-me1}$_2$, we employ
\begin{equation}\label{eq-me3}
\begin{split}
\int_{\Gamma_{cm}^\dlt} q_m^\dlt \nb^\mx \cdot  \wb^\dlt = -\int_{\Gamma_{cm}^\dlt} q_m^\dlt \nb^\cx \cdot  \wb^\dlt \\
= - \int_{\Om_c^\dlt} \nabla\cdot(\tilde q_m^\dlt \wb^\dlt) + \int_{\pd_\ext\Om_c^\dlt} \tilde q_m^\dlt \nb\cdot
\wb^\dlt\\
\rightarrow - \int_\Om \intY_{Z_c} (\nabla_x{q}^0 + \nabla_z{q}^1)\cdot\wb^0
- \int_\Om \intY_{Z_c} {q}^0\left( \nabla_x\cdot \wb^0 +  \nabla_z\cdot\wb^1\right)
+\int_{\pd_\ext\Om^\dlt} \bar\phi_c q^0 \nb\cdot\wb^0 \\
 = -\int_\Om \wb^0\cdot\intY_{Z_c} \nabla_z{q}^1 - \int_\Om{q}^0\intY_{\Gamma_Z}\wb^1\cdot\nb^\cx 
- \int_\Om \phi_c\nabla_x\cdot(q^0 \wb^0) + \int_{\pd\Om} \bar \phi_c q^0 \nb\cdot  \wb^0\;.
\end{split}
\end{equation}

In analogy we treat the interface integral in \eq{eq-me1}$_3$ (note $\vthetabf^\dlt = 0$ on $\pd_\ext\Om_c^\dlt$)
\begin{equation}\label{eq-me4}
\begin{split}
\int_{\Gamma_{cm}^\dlt} p_m^\dlt \nb^\cx \cdot \vthetabf^\dlt = \int_{\Om_c^\dlt} \nabla\cdot(\tilde p_m^\dlt \vthetabf^\dlt) - 
\int_{\pd_\ext\Om_c^\dlt} \tilde p_m^\dlt \nb\cdot\vthetabf^\dlt\\
\rightarrow \int_\Om \intY_{Z_c}(\nabla_x{p}^0 + \nabla_z{p}^1)\cdot\vthetabf^0 + 
 \int_\Om \intY_{Z_c}{p}^0 (\nabla_x\cdot\vthetabf^0 + \nabla_z\vthetabf^1)
 - \int_{\pd\Om} {p}^0  \bar\phi_c \nb\cdot\vthetabf^0\\
 = \int_\Om \vthetabf^0 \cdot \int_{\Gamma_Z} p^1 \nb^\cx + \int_\Om \intY_{Z_c}{p}^0 \nabla_z\cdot\vthetabf^1 + \int_\Om \phi_c \nabla_x\cdot(p^0\vthetabf^0) - \int_{\pd\Om} {p}^0  \bar\phi_c \nb\cdot\vthetabf^0
 \;.
\end{split}
\end{equation}
This can be rewritten further to get:
\begin{equation}\label{eq-me4a}
\begin{split}
\int_{\Gamma_{cm}^\dlt} p_m^\dlt \nb^\cx \cdot \vthetabf^\dlt 
\rightarrow \int_\Om \intY_{Z_c} \nabla_z{p}^1\cdot\vthetabf^0 + 
 \int_\Om \intY_{Z_c}{p}^0 \nabla_z\cdot\vthetabf^1 \\
+ \int_{\pd\Om} (\phi_c - \bar\phi_c) {p}^0 \nb\cdot\vthetabf^0 -\int_\Om {p}^0 \vthetabf^0 \nabla_x \phi_c\;.
\end{split}
\end{equation}
In many situations,  $\phi_c = \bar\phi_c$ and $\nabla_x \phi_c = 0$.

\subsection{Characteristic responses of the mesoscopic structure}
 Upon substituting the split \eq{eq-me11-i} in the local problem \eq{eq-me1} and using the linearity, problem \eq{eq-me15b-i} and the following three problems satisfied by the characteristic responses are obtained.
\begin{enumerate}

\item
Find $(\chibf^{ij},\eta^{ij}) \in \Hpdb(Z_c)/\RR^3\times L^2(Z_c)$, such that,
\begin{equation}\label{eq-me15a-i}
\begin{split}
2\eta_c \ipZc{\eebz{\chibf^{ij}}}{\eebz{\vthetabf}} - \ipZc{\eta^{ij}}{\nabla_z\cdot\vthetabf}
& = -2\eta \ipZc{\eebz{\tilde\omegabf^{ij} + \Pibf^{ij}}}{\eebz{\vthetabf}} \;,\\
\ipZc{\nabla_z\cdot\chibf^{ij}}{q}  & = - \ipZc{\nabla_z\cdot(\tilde\omegabf^{ij} + \Pibf^{ij})}{q}\;,
\end{split}
\end{equation}
for all $(\vthetabf,q) \in \Hpdb(Z_c)\times L^2(Z_c)$.

\item
Find $(\psibf^P,\hat \eta^P) \in \Hpdb(Z_c)/\RR^3\times L^2(Z_c)$, such that,
\begin{equation}\label{eq-me15c-i}
\begin{split}
2\eta_c \ipZc{\eebz{\psibf^P}}{\eebz{\vthetabf}} - \ipZc{\eta^P}{\nabla_z\cdot\vthetabf}
& = - \intY_{\Gamma_Z}\vthetabf\cdot\nb^\cx \;,\\
\ipZc{\nabla_z\cdot\psibf^P}{q} & =  0\;,
\end{split}
\end{equation}
for all $(\vthetabf,q) \in \Hpdb(Z_c)\times L^2(Z_c)$.

\item
Find $(\xibf^P,\hat \zeta^P) \in \Hpdb(Z_c)/\RR^3\times L^2(Z_c)$, such that,
\begin{equation}\label{eq-me15d-i}
\begin{split}
2\eta \ipZc{\eebz{\xibf^P}}{\eebz{\vthetabf}} - \ipZc{\hat\zeta^P}{\nabla_z\cdot\vthetabf}
& = - 2\eta_c \ipZc{\eebz{\tilde\omegabf^P}}{\eebz{\vthetabf}}\;,\\
\ipZc{\nabla_z\cdot\xibf^P}{q} & = - \ipZc{\nabla_z\cdot(\tilde\omegabf^P}{q}\;,
\end{split}
\end{equation}
for all $(\vthetabf,q) \in \Hpdb(Z_c)\times L^2(Z_c)$.

\end{enumerate}

\subsection{Hmogenized problem and homogenized coefficients of the double-porosity mdium} \label{apx-4} 
The first set of the HC given in \eq{eq-ma2-i} which are involved in \eq{eq-ma10-i}$_1$ can be identified easily from the \lhs of \eq{eq-ma1-i}$_1$, 
where the split \eq{eq-me11-i} is substituted,
\begin{equation}\label{eq-apx1}
\begin{split}
\int_\Om \left[\aYm{\omegabf^{kl}+\Pibf^{kl}}{\Pibf^{ij}}e_{kl}^x(\ub^0) - \aYm{p^0}{\Pibf^{ij}}\right]e_{ij}^x(\vb^0)\\
\int_\Om \vb^0\intY_{\Gamma_Z}\nb^\mx\cdot[\pi^k (\pd_k^x p^0 - f_k^f) + \phi^i w_i^0] - \int_\Om \phi_c \nabla_x\cdot(p^0\vb^0) 
+ \int_{\pd\Om } \bar \phi_c p^0\nb\cdot\vb^0 = \dots \;.
\end{split}
\end{equation}
Now \eq{eq-ma10-i}$_1$ is obtained upon substituting the HC expressed in \eq{eq-ma2-i}; recall  \eq{eq-ma6-i} and \eq{eq-ma7-i} which allows to introduce the bulk force $\fb^\blk$.
In analogy,  the \lhs of \eq{eq-ma1-i}$_2$ yields
\begin{equation}\label{eq-apx2}
\begin{split}
\int_\Om \left[\bYm{q^0}{\omegabf^{kl}+\Pibf^{kl}}e_{kl}^x(\dot\ub^0) - \bYm{q^0}{\omegabf^P}\dot p^0\right]\\
+
\int_\Om \pd_j^x q^0[\cYm{z_k+\pi^k}{z_j} (\pd_k^x p^0 - f_k^f) + \cYm{z_j}{\vphi^k} w_k^0] + \int_\Om  q^0 \dot p^0 \intY_{Z_m} M  \\
 - \int_\Om \phi_c \nabla_x\cdot(q^0\wb^0) -\int_\Om q^0 \intY_{\Gamma_Z}\nb^\cx\cdot\wb^1 = \dots\;.
\end{split}
\end{equation}
Rather than the direct substitution of $\wb^1$ using the split \eq{eq-me11-i}, the last intergral can be expressed using the local mass conservation \eq{eq-me10} emloyed with $\gamma=0$,
\begin{equation}\label{eq-apx3}
\begin{split}
-\intY_{\Gamma_Z}\nb^\cx \cdot \wb^1 =  \intY_{\Gamma_Z}\nb^\cx \cdot \dot{\tilde\ub}^1 + \phi_c \nabla_x\cdot(\dot{\ub}^0 +\wb^0)\;,\quad \mbox{ a.e. in }\Om \;.
\end{split}
\end{equation}
Now, using the split, the last two integrals in \eq{eq-apx2} yield
\begin{equation}\label{eq-apx4}
\begin{split}
 - \int_\Om \phi_c \nabla_x\cdot(q^0\wb^0) + \int_\Om \left[
\phi_c \nabla_x\cdot(\wb^0 + \dot\ub^0) + \intY_{\Gamma_Z}\nb^\cx \cdot \dot{\tilde\ub}^1
\right]\\
= - \int_\Om \phi_c\wb^0\cdot\nabla_x q^0 +  \int_\Om q^0 \left[ \phi_c\nabla_x\cdot\dot{\ub}^0 
+ \intY_{\Gamma_Z}\nb^\cx \cdot (\omegabf^{ij}e_{ij}^x(\dot{\ub}^0) +\omegabf^P \dot p^0 )\right]\;.
\end{split}
\end{equation}
Eg. \eq{eq-ma10-i}$_2$ is now obtained using HC \eq{eq-ma3-i} substituted in the above expressions, whereby \eq{eq-ma6-i} is employed.

To derive \eq{eq-ma10-i}$_3$, we substitute the split \eq{eq-me11-i} in the \lhs of \eq{eq-ma1-i}$_3$. Due to Proposition~\ref{prop-1},(iii) and (iv),
all terms involving $p^0$ and $\dot p^0$ vanish, indeed
\begin{equation}\label{eq-apx5}
\begin{split}
\left(2\eta \ipZc{\eebz{\psibf^P}}{\eebz{\Pibf^{ij}}} - \ipZc{\hat\eta^P}{\delta_{ij}} \right) \pd_j^x \vtheta_i^0 & = 0\;,\\
\left(2\eta \ipZc{\eebz{\xibf^P + \tilde\omegabf^P}}{\eebz{\Pibf^{ij}}} - \ipZc{\hat\zeta^P}{\delta_{ij}} \right) \pd_j^x \vtheta_i^0 & = 0\;.
\end{split}
\end{equation}
All terms involving $\eeb{\wb^0}$ are expressed by $\eeb{\vthetabf}:\Scalbf\eeb{\wb^0}$, where $\Scalbf$ is given in \eq{eq-ma4-i}$_1$. Due to Proposition~\ref{prop-1},(i) and (ii), the same tensor is collects contributions involving $\eeb{\dot\ub^0}$, as follows from
\begin{equation}\label{eq-apx6}
\begin{split}
e_{ij}^x(\vthetabf^0)\left(2\eta \ipZc{\eebz{\psibf^{kl}}}{\eebz{\Pibf^{ij}}} - \ipZc{\hat\vphi^{kl}}{\nabla_z\cdot\Pibf^{ij}}\right)e_{kl}^x(\wb^0 + \dot\ub^0)\;.
\end{split}
\end{equation}
Further, it is easy to see that (in analogy with \eq{eq-ma7-i}, the term involving $\fb^f$ is moved form the \rhs to the \lhs in \eq{eq-ma1-i}$_3$), 
\begin{equation}\label{eq-apx7}
\begin{split}
\int_\Om \vthetabf^0 \cdot \int_{\Gamma_Z} p^1 \nb^\cx + \int_\Om \phi_c(\nabla_xp^0 - \fb^f)\cdot\vthetabf^0 
= \int_\Om \vthetabf^0 \cdot( \Pcalbf (\nabla_xp^0 - \fb^f) + \Hcalbf \wb^0)\;.
\end{split}
\end{equation}
We conclude \eq{eq-ma10-i}$_3$ is obtained from \eq{eq-ma10-i}$_3$ using HC \eq{eq-ma4-i} and \eq{eq-ma6-i}.

\end{appendices}

\bibliographystyle{plain}      


\begin{thebibliography}{10}

\bibitem{Allaire1992}
G.~Allaire.
\newblock Homogenization and two-scale convergence.
\newblock {\em SIAM Journal on Mathematical Analysis}, 23(6):1482–1518, 1992.

\bibitem{Arbogast1990}
T.~Arbogast, J.~Douglas, and U.~Hornung.
\newblock Derivation of the double porosity model of single phase flow via
  homogenization theory.
\newblock {\em SIAM Journal on Mathematical Analysis}, 21(4):823--836, 1990.

\bibitem{Auriault-validity-Brinkman2009}
J.L. Auriault.
\newblock On the domain of validity of {B}rinkman’s equation.
\newblock {\em Transport in porous media}, 79(2):215--223, 2009.

\bibitem{Auriault1992}
J.L. Auriault and C.~Boutin.
\newblock Deformable porous media with double porosity. quasi-statics. i:
  Coupling effects.
\newblock {\em Transport in porous media}, 7(1):63--82, 1992.

\bibitem{Auriault1993}
J.L. Auriault and C.~Boutin.
\newblock Deformable porous media with double porosity. quasi-statics. ii:
  Memory effects.
\newblock {\em Transport in porous media}, 10(2):153--169, 1993.

\bibitem{Auriault1977}
J.L. Auriault and E.~Sanchez-Palencia.
\newblock Etude du comportment macroscopique d'un milieu poreux satur\'e
  d\'eformable.
\newblock {\em J. de M\'ecanique}, 16(4):575--603, 1977.

\bibitem{Barenblatt1960}
G.I. Barenblatt, I.P. Zheltov, and I.N. Kochina.
\newblock Basic concepts in the theory of seepage of homogeneous liquids in
  fissured rocks.
\newblock {\em Journal of Applied Mathematics and Mechanics},
  24(5):1286–1303, 1960.

\bibitem{Bensoussan1978book}
A.~Bensoussan, J.L. Lions, and G.~Papanicolaou.
\newblock {\em Asymptotic methods in periodic media}.
\newblock North-Holland, 1978.

\bibitem{Berryman2002}
J.G. Berryman.
\newblock Models for computing geomechanical constants of double-porosity
  materials from the constituents properties.
\newblock {\em Journal of Geophysical Research}, 107(B3), 2002.

\bibitem{Berryman2012}
J.G. Berryman.
\newblock Poroelastic response of orthotropic fractured porous media.
\newblock {\em Transport in Porous Media}, 93(2):293–307, 2011.

\bibitem{Biot1941}
M.A. Biot.
\newblock General theory of three-dimensional consolidation.
\newblock {\em Journal of Applied Physics}, 12(2):155–164, 1941.

\bibitem{Biot1955}
M.A. Biot.
\newblock Theory of elasticity and consolidation for a porous anisotropic
  solid.
\newblock {\em Journal of Applied Physics}, 26(2):182–185, 1955.

\bibitem{Brinkman1947}
H.C. Brinkman.
\newblock On the permeability of media consisting of closely packed porous
  particles.
\newblock {\em Applied Scientific Research}, 1(1):81--86, 1947.

\bibitem{Brinkman1949}
H.C. Brinkman.
\newblock A calculation of the viscous force exerted by a flowing fluid on a
  dense swarm of particles.
\newblock {\em Applied Scientific Research}, 1(1):27--34, 1949.

\bibitem{Brown2015}
D.L. Brown, Y.~Efendiev, G.~Li, and V.~Savatorova.
\newblock Homogenization of high-contrast brinkman flows.
\newblock {\em Multiscale Modeling \& Simulation}, 13(2):472--490, 2015.

\bibitem{Chandesris-TiPM2009}
M.~Chandesris and D.~Jamet.
\newblock Jump conditions and surface-excess quantities at a fluid/porous
  interface: A multi-scale approach.
\newblock {\em Transport in Porous Media}, 78(3):419--438, 2009.

\bibitem{cimrman_2014:sfepy}
R.~Cimrman.
\newblock {SfePy} - write your own {FE} application.
\newblock In Pierre de~Buyl and Nelle Varoquaux, editors, {\em Proceedings of
  the 6th European Conference on Python in Science (EuroSciPy 2013)}, pages
  65--70, 2014.

\bibitem{Cioranescu-etal-2008}
D.~Cioranescu, A.~Damlamian, and G.~Griso.
\newblock The periodic unfolding method in homogenization.
\newblock {\em SIAM Journal on Mathematical Analysis}, 40(4):1585--1620, 2008.

\bibitem{CDG-Stokes-2005}
D.~Cioranescu, A.~Damlamian, G.~Griso, and et~al.
\newblock The {S}tokes problem in perforated domains by the periodic unfolding
  method. new trends in continuum mechanics.
\newblock {\em Theta Series in Advanced Mathematics}, 3:67--80, 2005.

\bibitem{Gailani2011}
G.~Gailani and S.~Cowin.
\newblock Ramp loading in russian doll poroelasticity.
\newblock {\em Journal of the Mechanics and Physics of Solids}, 59(1):103--120,
  2011.

\bibitem{Hornung1997book}
U.~Hornung.
\newblock {\em Homogenization and porous media}.
\newblock Springer, Berlin, 1997.

\bibitem{Jaeger-Mikelic-2009}
W.~J{\"a}ger and A.~Mikeli{\'c}.
\newblock Modeling effective interface laws for transport phenomena between an
  unconfined fluid and a porous medium using homogenization.
\newblock {\em Transport in porous media}, 78(3):489--508, 2009.

\bibitem{Jamet-TiPM2009}
D.~Jamet, M.~Chandesris, and B.~Goyeau.
\newblock On the equivalence of the discontinuous one- and two-domain
  approaches for the modeling of transport phenomena at a fluid/porous
  interface.
\newblock {\em Transport in Porous Media}, 78(3):403--418, 2009.

\bibitem{Lesinigo-etal-multi-Brinkman2011}
M.~Lesinigo, C.~D’Angelo, and A.~Quarteroni.
\newblock A multiscale {D}arcy--{B}rinkman model for fluid flow in fractured
  porous media.
\newblock {\em Numerische Mathematik}, 117(4):717--752, 2011.

\bibitem{Czochra-Mikelic-MMS-2012}
A.~Marciniak-Czochra and A.~Mikelic.
\newblock Effective pressure interface law for transport phenomena between an
  unconfined fluid and a porous medium using homogenization.
\newblock {\em SIAM, Multiscale Modeling and Simulation}, 10(2):285--305, 2012.

\bibitem{OCHOATAPIA19952647}
A.~Ochoa-Tapia and S.~Whitaker.
\newblock Momentum transfer at the boundary between a porous medium and a
  homogeneous fluid—ii. comparison with experiment.
\newblock {\em International Journal of Heat and Mass Transfer},
  38(14):2647--2655, 1995.

\bibitem{OchoaTapia19952635}
J.~Ochoa-Tapia and S.~Whitaker.
\newblock Momentum transfer at the boundary between a porous medium and a
  homogeneous fluid—i. theoretical development.
\newblock {\em International Journal of Heat and Mass Transfer},
  38(14):2635--2646, 1995.

\bibitem{Rohan-AMC}
E.~Rohan and V.~Lukeš.
\newblock Modeling nonlinear phenomena in deforming fluid-saturated porous
  media using homogenization and sensitivity analysis concepts.
\newblock {\em Applied Mathematics and Computation}, 267:583–595, 2015.

\bibitem{rohan-etal-jmps2012-bone}
E.~Rohan, S.~Naili, R.~Cimrman, and T.~Lemaire.
\newblock Multiscale modeling of a fluid saturated medium with double porosity:
  Relevance to the compact bone.
\newblock {\em Journal of the Mechanics and Physics of Solids}, 60(5):857--881,
  2012.

\bibitem{rohan-etal-CMAT2015}
E.~Rohan, S.~Naili, and T.~Lemaire.
\newblock Double porosity in fluid-saturated elastic media: deriving effective
  parameters by hierarchical homogenization of static problem.
\newblock {\em Continuum Mechanics and Thermodynamics}, 28(5):1263–1293, Apr
  2015.

\bibitem{RTL-CC2015}
E.~Rohan, J.~Turjanicov\'{a}, and V.~Luke\v{s}.
\newblock Modelling flows in multi-porous media using homogenization with
  application to liver lobe perfusion.
\newblock In J.~Kruis, Y.~Tsompanakis, and B.H.V. Topping, editors, {\em
  Proceedings of the Fifteenth International Conference on Civil, Structural
  and Environmental Engineering Computing}. Civil-Comp Press, Stirlingshire,
  UK, Paper 148, 2015. doi:10.4203/ccp.108.148.

\bibitem{Rohan-etal-CaS2017}
E.~Rohan, J.~Turjanicová, and V.~Lukeš.
\newblock A darcy-brinkman model of flow in double porous media -- two-level
  homogenization and computational modelling.
\newblock {\em Computers \& Structures}, 207:95–110, 2018.

\bibitem{Royer-Auriault-Boutin-1996}
P.~Royer, J.L. Auriault, and C.~Boutin.
\newblock Macroscopic modeling of double-porosity reservoirs.
\newblock {\em Journal of Petroleum Science and Engineering}, 16(4):187--202,
  1996.

\bibitem{Saffman1971}
P.G. Saffman.
\newblock On the boundary condition at the surface of a porous medium.
\newblock {\em Studies in Applied Mathematics}, 50(2):93--101, 1971.

\bibitem{Sanchez1980Book}
E.~Sanchez-Palencia.
\newblock {\em Non-homogeneous media and vibration theory}.
\newblock Number 127 in Lecture Notes in Physics. Springer, Berlin, 1980.

\bibitem{Sandstrom-Larssen-CMAME2016}
C.~Sandström, F.~Larsson, and K.~Runesson.
\newblock Homogenization of coupled flow and deformation in a porous material.
\newblock {\em Computer Methods in Applied Mechanics and Engineering},
  308:535–551, 2016.

\bibitem{trucu2012three}
D.~Trucu, M.A.J. Chaplain, and A.~Marciniak-Czochra.
\newblock Three-scale convergence for processes in heterogeneous media.
\newblock {\em Applicable Analysis}, 91(7):1351--1373, 2012.

\end{thebibliography}

\end{document}